\documentclass[preprintnumbers]{revtex4}
\UseRawInputEncoding
\usepackage{amssymb}
\usepackage{amsmath}
\usepackage{graphicx}
\usepackage{calligra}
\usepackage{mathrsfs}
\usepackage{dcolumn}
\usepackage{bm}
\usepackage{subfigure}
\usepackage{color}
\usepackage{CJKutf8}
\usepackage{float}
\usepackage{multirow} 
\usepackage{booktabs} 
\usepackage{array}    
\usepackage{caption}  
\numberwithin{equation}{section}
\numberwithin{figure}{section}

\begin{document}
\title{High-order QED correction impacts on phase transition of the Euler-Heisenberg dS spacetime}
\author{Shan-Xia Bao$^{1,2}$, Meng-Sen Ma$^{1,2}$, Huai-Fan Li$^{3,4}$, Yun-Zhi Du$^{1,2,3}$}
\thanks{\emph{e-mail:duyzh22@sxdtdx.edu.cn }(corresponding author)}

\address{$^1$Department of Physics, Shanxi Datong University, Datong 037009, China\\
$^2$Institute of Theoretical Physics, Shanxi Datong University, Datong, 037009, China\\
$^3$State Key Laboratory of Quantum Optics and Quantum Optics Devices, Shanxi University, Taiyuan 030006, China\\
$^4$College of General Education, Shanxi College of Technology, Shuozhou 036000, China\\}

\begin{abstract}

Recent studies have demonstrated that AdS black holes possess the basic characteristics of a standard thermodynamic system. Concurrently, the thermodynamic properties of spacetimes featuring multiple horizons with distinct radiation temperatures have also attracted research interest. In this work, considering the high-order quantum electrodynamics (QED) correction, we initially establish an equivalent thermodynamic system for the coexistence region of black hole and cosmological horizons. On this basis, we conduct a detailed investigation into the thermodynamic properties of this dual-horizon coexistence region. Our results demonstrate that this equivalent thermodynamic system exhibits van der Waals-like thermodynamic behavior. Furthermore, we introduce a nonlinear parameter $\gamma$ to analyze its impact on phase transitions within the equivalent thermodynamic system. Under specific conditions, the system undergoes first- or second-order phase transitions for $\gamma=0$, and zeroth- or second-order phase transitions for $\gamma\neq0$. Finally, by utilizing the generalized off-shell Helmholtz free energy within the thermodynamic topological framework for black holes, we extend this methodology to investigate the topological properties of de Sitter (dS) spacetime. We compute the topological number characterizing the coexistence region of dual horizons in Euler-Heisenberg (EH) dS spacetime using equivalent thermodynamic state parameters. Additionally, we investigate the influence of the nonlinear parameter $\gamma$ on the thermodynamic characteristics of the equivalent system.

\end{abstract}

\maketitle

\section{Introduction}\label{one}

Black hole physics represents an interdisciplinary field intersecting general relativity, quantum mechanics, thermodynamics and statistical physics, particle physics, and string theory. Research exploring the connection between the classical thermodynamic quantities of black holes and quantum gravity plays a crucial role in modern physics. Consequently, investigations into both the classical thermodynamic behavior and the microscopic structure of black holes offer an essential avenue toward understanding quantum gravity.

In recent years, researches into the thermodynamic properties and internal microstructure of black holes have advanced significantly from diverse perspectives. Initial studies of phase transitions between large and small black holes in spacetimes with a negative cosmological constant revealed that AdS black holes exhibit van der Waals-like phase transitions \cite{1,2,3,4,5,6,7,8,9,10,11,12,13,14,15,16,17,18,19}. This was followed by growing interest in applying topological methods to black hole thermodynamics \cite{20,21}. Notably, investigations of topological numbers associated with various AdS black holes have identified distinct topological classes \cite{22,23,24,25,26,27,28,29}. However, research on black hole thermodynamics remains predominantly focused on AdS spacetimes. Studies addressing thermodynamics in dS spacetime have, thus far, largely adopted the approach of treating the different horizons as independent thermodynamic systems \cite{30,31,32}. This approach is fundamentally limited because the thermodynamic state parameters associated with each horizon in dS spacetime are interdependent. Since these parameters are functions of the fundamental spacetime parameters-such as mass $(M)$, charge $(Q)$, and angular momentum $(J)$-they are inherently interrelated. Consequently, analyzing the horizons in isolation yields incomplete conclusions.Recently, to analyze the thermodynamic properties of the region in de Sitter (dS) spacetime where both a black hole horizon and a cosmological horizon coexist, we considered the inherent correlation between the two horizons \cite{33,34,35}. Building on this, we treated the combined thermodynamic system formed by these two horizons as a unified equivalent thermodynamic system. This equivalent system simultaneously satisfy the universal first law of thermodynamics and the boundary conditions specific to the coexistence region. Constrained by these principles, we derived the thermodynamic quantities for the equivalent system in the coexistence region. Crucially, we found that the specific heat capacity ${{C}_{Q,l}}$  of this region exhibits a characteristic Schottky-type dependence on the equivalent temperature \cite{36,37}. This finding establishes a foundation for investigating the thermodynamic properties of the two-horizon coexistence region.

Based on Dirac's positron theory, Euler and Heisenberg developed a novel framework for describing electromagnetic fields. By modifying Maxwell's equations in vacuum \cite{38}, they derived an effective Lagrangian density for the electromagnetic field. This Lagrangian incorporates higher-order nonlinear electromagnetic (NEM) field terms \cite{39}. Within quantum electrodynamics (QED), Schwinger subsequently reformulated this non-perturbative, one-loop effective Lagrangian density, which captures the essential characteristics of the Euler-Heisenberg (EH) NEM field \cite{40}. Notably, when the electric field strength exceeds a critical value (${{m}^{2}}{{c}^{3}}/e\hbar $ ), QED effects induce vacuum pair production \cite{41}. Coupling this one-loop effective Lagrangian density to the Einstein field equations, Yajima $et~al$. derived the corresponding Euler-Heisenberg black hole solution \cite{42}. Subsequent researches on the thermodynamic properties of the EH AdS black hole also yielded significant conclusions \cite{43,44,45}.

This study extends the methodology for investigating thermodynamic properties of AdS black holes to the coexistence region of black hole and cosmological horizons. First, leveraging analytical approaches commonly applied to conventional thermodynamic systems and AdS black holes, we examine zeroth-order, first-order and second-order phase transitions in the equivalent thermodynamic system, this analysis identifies the key factors governing its phase behavior. Subsequently, we explore the topological properties of this equivalent thermodynamic system, comparing and contrasting them with those of AdS black holes. Through multi-perspective investigations into the thermodynamics of the dual-horizon region, our work deepens the understanding of dS spacetime. Our findings establish a novel pathway for probing the thermodynamics of dS spacetime and provide fresh perspectives for revealing the nature of microscopic particles within black holes.

This paper is structured as follows. In Sec.2, we briefly review the equivalent thermodynamic quantities of EH dS spacetime to establish foundational concepts. In Sec.3, we analyze critical behavior and phase transitions in the dual-horizon thermodynamic system, with emphasis on the impact of nonlinear parameters. In Sec.4, we extend topological analysis methods from AdS black holes to the equivalent thermodynamic system of dS spacetime's dual-horizon region, comparing topological characteristics between both systems. Concluding remarks are presented in Sec.5.

\section{Euler-Heisenberg de Sitter spacetime} \label{two}

The static spherically symmetric black hole solution in a topological spacetime with non-linear source were given as \cite{46,47,48,49}:
\begin{eqnarray}
d{{s}^{2}}=-g(r)d{{t}^{2}}+{{g}^{-1}}d{{r}^{2}}+{{r}^{2}}d\Omega _{2}^{2} \label{2.1}
\end{eqnarray}
With the horizon function:
\begin{eqnarray}
g(r)=k-\frac{2M}{r}+\frac{{{Q}^{2}}}{{{r}^{2}}}-\frac{{{r}^{2}}}{{{l}^{2}}}-\frac{2{{Q}^{4}}\alpha }{5{{r}^{6}}} \label{2.2}
\end{eqnarray}
Here $M$ and $Q$ are the black hole mass and charge, and the last term in Eq. (2.2) indicates the effect of the non-linearity. In the following we mainly focus on the solution $\Lambda>0$ with $k=1$ , i.e., the de-Sitter spacetime with a black hole. In this system there are two horizons, one is of dS black hole $r_{+}$, another is of cosmology $r_{c}$. And these two horizons are satisfied with the expression $g({{r}_{+,c}})=0$.

Based on $g({{r}_{+,c}})=0$, we have:
\begin{eqnarray}
M=\frac{r}{2}\left( 1+\frac{{{Q}^{2}}}{{{r}^{2}}}-\frac{{{r}^{2}}}{{{l}^{2}}}-\frac{2{{Q}^{4}}\alpha }{5{{r}^{6}}} \right) \label{2.3}
\end{eqnarray}
\begin{eqnarray}
M=\frac{{{r}_{c}}x(1+x)}{2(1+x+{{x}^{2}})}+\frac{{{Q}^{2}}(1+x)(1+{{x}^{2}})}{2{{r}_{c}}x(1+x+{{x}^{2}})}-\frac{{{Q}^{4}}\alpha (1-{{x}^{8}})}{5r_{c}^{5}{{x}^{5}}(1-{{x}^{3}})} \label{2.4}
\end{eqnarray}
\begin{eqnarray}
\frac{1}{{{l}^{2}}}=\frac{1}{r_{c}^{2}(1+x+{{x}^{2}})}-\frac{{{Q}^{2}}}{r_{c}^{4}x(1+x+{{x}^{2}})}+\frac{2{{Q}^{4}}\alpha (1-{{x}^{5}})}{5r_{c}^{8}{{x}^{5}}(1-{{x}^{3}})} \label{2.5}
\end{eqnarray}
where $x={{r}_{+}}/{{r}_{c}}$ is the ratio of the black hole horizon location to the cosmological horizon location.

The radiation temperature $T_{+}$ associated with the black hole horizon is given by:
\begin{eqnarray}
{{T}_{+}}&=&\frac{g'({{r}_{+}})}{4\pi }=\frac{1}{4\pi }\left( \frac{k}{{{r}_{+}}}-\frac{3{{r}_{+}}}{{{l}^{2}}}-\frac{{{Q}^{2}}}{r_{+}^{3}}+\frac{2{{Q}^{4}}\alpha }{r_{+}^{7}} \right) \notag \\
&=&\frac{(1-x)}{4\pi {{r}_{c}}x(1+x+{{x}^{2}})}\left( k(1+2x)-\frac{{{Q}^{2}}(1+2x+3{{x}^{2}})}{r_{c}^{2}{{x}^{2}}}+\frac{2{{Q}^{2}}\gamma [5(1-{{x}^{3}})-3{{x}^{3}}(1-{{x}^{5}})]}{5r_{c}^{2}{{x}^{6}}{{(1-x)}^{2}}} \right) \label{2.6}
\end{eqnarray}
here $\gamma =\frac{{{Q}^{2}}\alpha }{r_{c}^{4}}$ is nonlinear parameters.

The radiation temperature $T_{c}$ associated with the cosmological horizon is given by:
\begin{eqnarray}
{{T}_{c}}&=&-\frac{g'({{r}_{c}})}{4\pi }=-\frac{1}{4\pi }\left( \frac{k}{{{r}_{c}}}-\frac{3{{r}_{c}}}{{{l}^{2}}}-\frac{{{Q}^{2}}}{r_{c}^{3}}+\frac{2{{Q}^{4}}\alpha }{r_{c}^{7}} \right) \notag \\
&=&\frac{(1-x)}{4\pi {{r}_{c}}(1+x+{{x}^{2}})}\left( k(2+x)-\frac{{{Q}^{2}}(3+2x+{{x}^{2}})}{r_{c}^{2}x} \right)+\frac{2{{Q}^{2}}\gamma [3(1-{{x}^{5}})-5{{x}^{5}}(1-{{x}^{3}})]}{20\pi r_{c}^{3}{{x}^{5}}(1-{{x}^{3}})} \label{2.7}
\end{eqnarray}

From Eq. (2.6), it follows that $T_{+}=0$ when the electrostatic potential on the black hole horizon satisfies:
\begin{eqnarray}
\frac{{{Q}^{2}}}{r_{A}^{2}}=k\frac{1+2x}{1+2x+3{{x}^{2}}}+\frac{2{{Q}^{4}}\alpha (5-8{{x}^{3}}+3{{x}^{8}})}{5r_{c}^{6}{{x}^{6}}{{(1-x)}^{2}}(1+2x+3{{x}^{2}})} \label{2.8}
\end{eqnarray}
Then the radiation temperature $T_{c}$ of the cosmological horizon corresponding to the black hole horizon position where $T_{+}=0$ is given by:
\begin{eqnarray}
{{T}_{c}}({{T}_{+}}=0)=\frac{k(1+x){{(1-x)}^{2}}}{2\pi {{r}_{c}}(1+2x+3{{x}^{2}})}-\frac{4{{Q}^{4}}\alpha (1+x)(1+{{x}^{2}}){{(1-x)}^{2}}(3+4x+3{{x}^{2}})}{5\pi r_{c}^{7}{{x}^{5}}(1+2x+3{{x}^{2}})} \label{2.9}
\end{eqnarray}
Subject to the boundary conditions (2.9), the universal first law of thermodynamics takes the form:
\begin{eqnarray}
dM={{T}_{eff}}dS-{{P}_{eff}}dV+{{\varphi }_{eff}}dQ \label{2.10}
\end{eqnarray}
Then, we have \cite{33,34,35}:
\begin{eqnarray}
{{P}_{eff}}&=&-\frac{(1-x)}{8\pi r_{c}^{2}{{x}^{5}}}\left\{ k\left( x(1+x)F'(x)/2-\frac{(1+2x)}{(1+x+{{x}^{2}})}F(x) \right) \right. \notag \\
&&+\frac{{{Q}^{2}}\gamma }{r_{c}^{2}{{x}^{6}}}\left( x(1+x)(1+{{x}^{2}})(1+{{x}^{4}})F'(x)/2-\frac{[8{{x}^{8}}(1-{{x}^{3}})+(1-{{x}^{8}})(5-8{{x}^{3}})]}{5(1-{{x}^{3}})(1-x)}F(x) \right) \notag \\
&&-\left. \frac{{{Q}^{2}}}{r_{c}^{2}{{x}^{2}}}\left( x(1+x)(1+{{x}^{2}})F'(x)/2-\frac{(1+2x+3{{x}^{2}})}{(1+x+{{x}^{2}})}F(x) \right) \right\} \label{2.11}
\end{eqnarray}
\begin{eqnarray}
{{T}_{eff}}&=&\frac{1-x}{4\pi {{r}_{c}}{{x}^{5}}}\left\{ \left[ (1+x)(1+{{x}^{3}})-2{{x}^{2}} \right] \right.-\frac{{{Q}^{2}}}{r_{c}^{2}{{x}^{2}}}\left[ (1+x+{{x}^{2}})(1+{{x}^{4}})-2{{x}^{3}} \right] \notag \\
&&+\left. \frac{{{Q}^{2}}\gamma (5-3{{x}^{3}}+3{{x}^{8}}-5{{x}^{11}})}{5r_{c}^{2}{{x}^{11}}(1-x)} \right\} \label{2.12}
\end{eqnarray}
where:
\begin{eqnarray}
&&F(x)=\frac{8}{5}{{(1-{{x}^{3}})}^{2/3}}+\frac{2}{5(1-{{x}^{3}})}-1=1+{{x}^{2}}+{{f}_{0}}(x) \notag \\
&&{{f}_{0}}(x)=\frac{8}{5}{{(1-{{x}^{3}})}^{2/3}}-\frac{8+5{{x}^{2}}-10{{x}^{3}}-5{{x}^{5}}}{5(1-{{x}^{3}})} \label{2.13}
\end{eqnarray}
\begin{eqnarray}
&&V={{V}_{c}}-{{V}_{+}}=\frac{4\pi }{3}r_{c}^{3}(1-{{x}^{3}}), ~~S=\pi r_{c}^{2}F(x) \notag \\
&&{{\varphi }_{eff}}=\frac{Q}{{{r}_{c}}x}\left( \frac{(1+x)(1+{{x}^{2}})}{(1+x+{{x}^{2}})}-\frac{4\gamma (1-{{x}^{8}})}{5{{x}^{4}}(1-{{x}^{3}})} \right) \label{2.14}
\end{eqnarray}
\section{Thermodynamic properties of the EH dS spacetime} \label{three}

Black hole phase transitions serve as a bridge connecting general relativity, thermodynamics, and quantum physics. They reveal the essence of black holes as complex thermodynamic systems and provide a new research perspective for quantum gravity. This section investigates the phase transitions in the equivalent thermodynamic system within the region of coexisting horizons in EH-dS spacetime. First, we employ Eqs. (2.11)-(2.14) to generate the $P_{eff}-V$ curves at constant temperature, as shown in FIG. 3.1.

\begin{figure}[htbp]
\centering
\subfigure[]{\includegraphics[width=8cm,height=4.5cm]{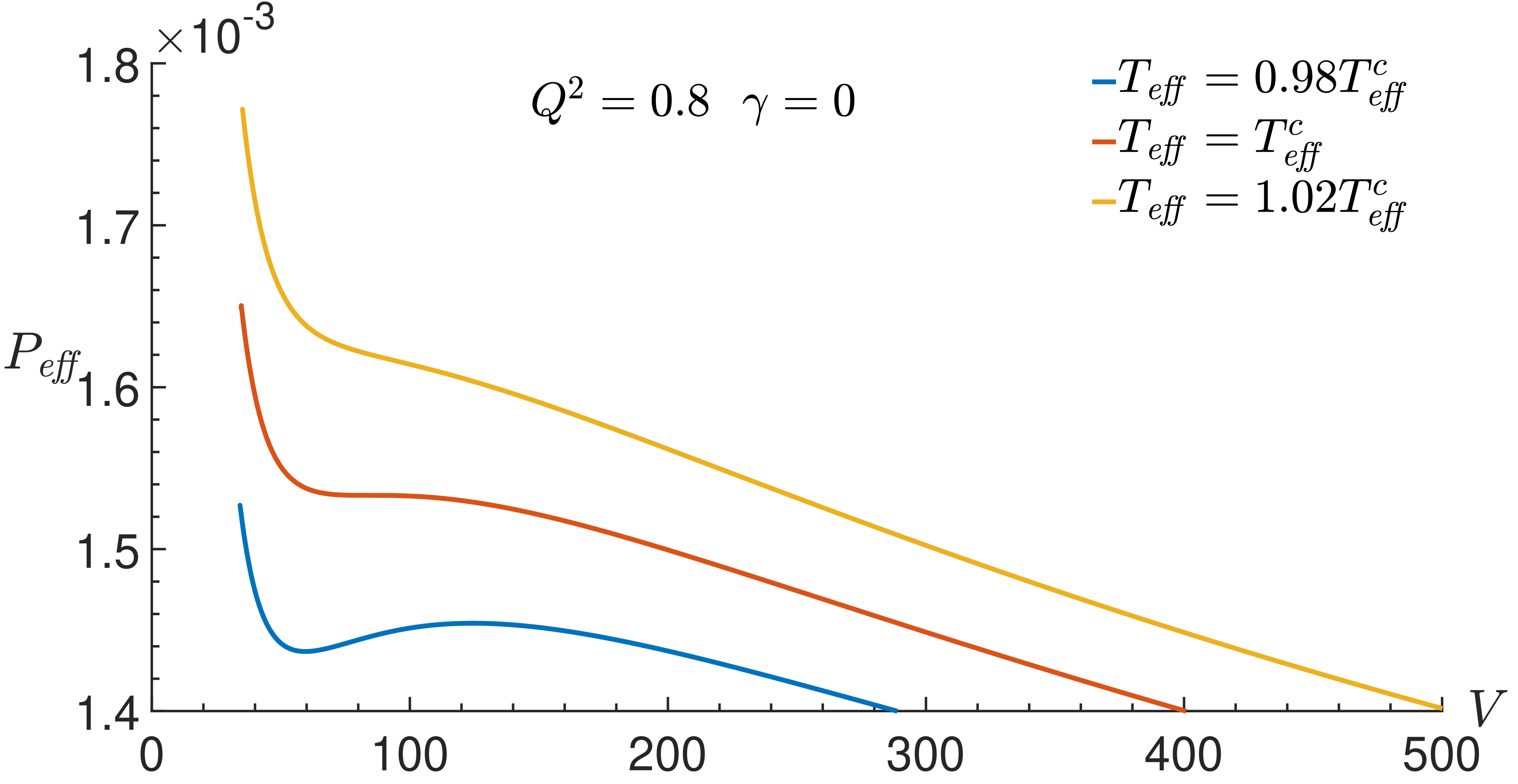}\label{Fig1a}}~
\hspace{0.000001cm}
\subfigure[]{\includegraphics[width=8cm,height=4.5cm]{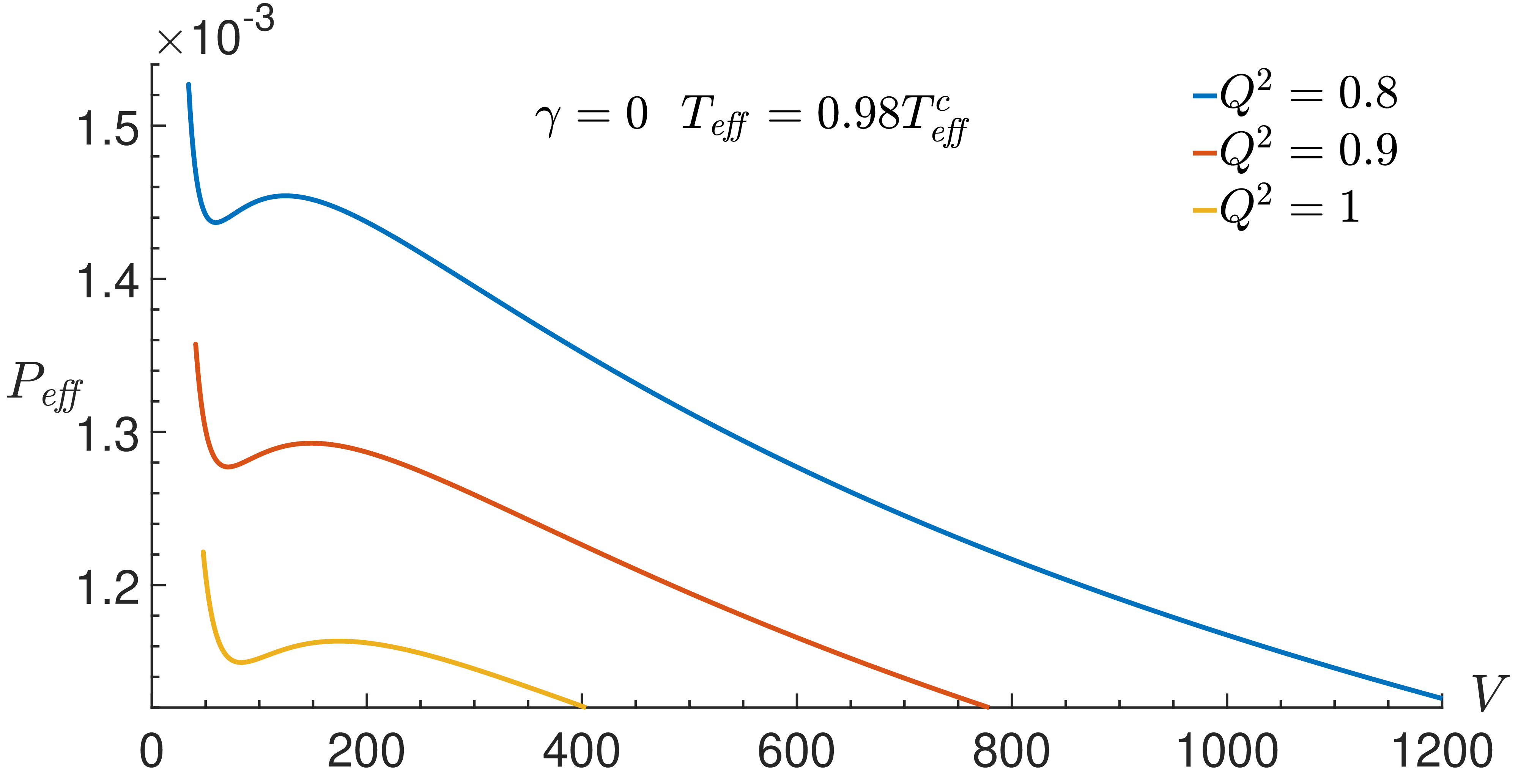}\label{Fig1b}}~
\\
\subfigure[]{\includegraphics[width=8cm,height=4.5cm]{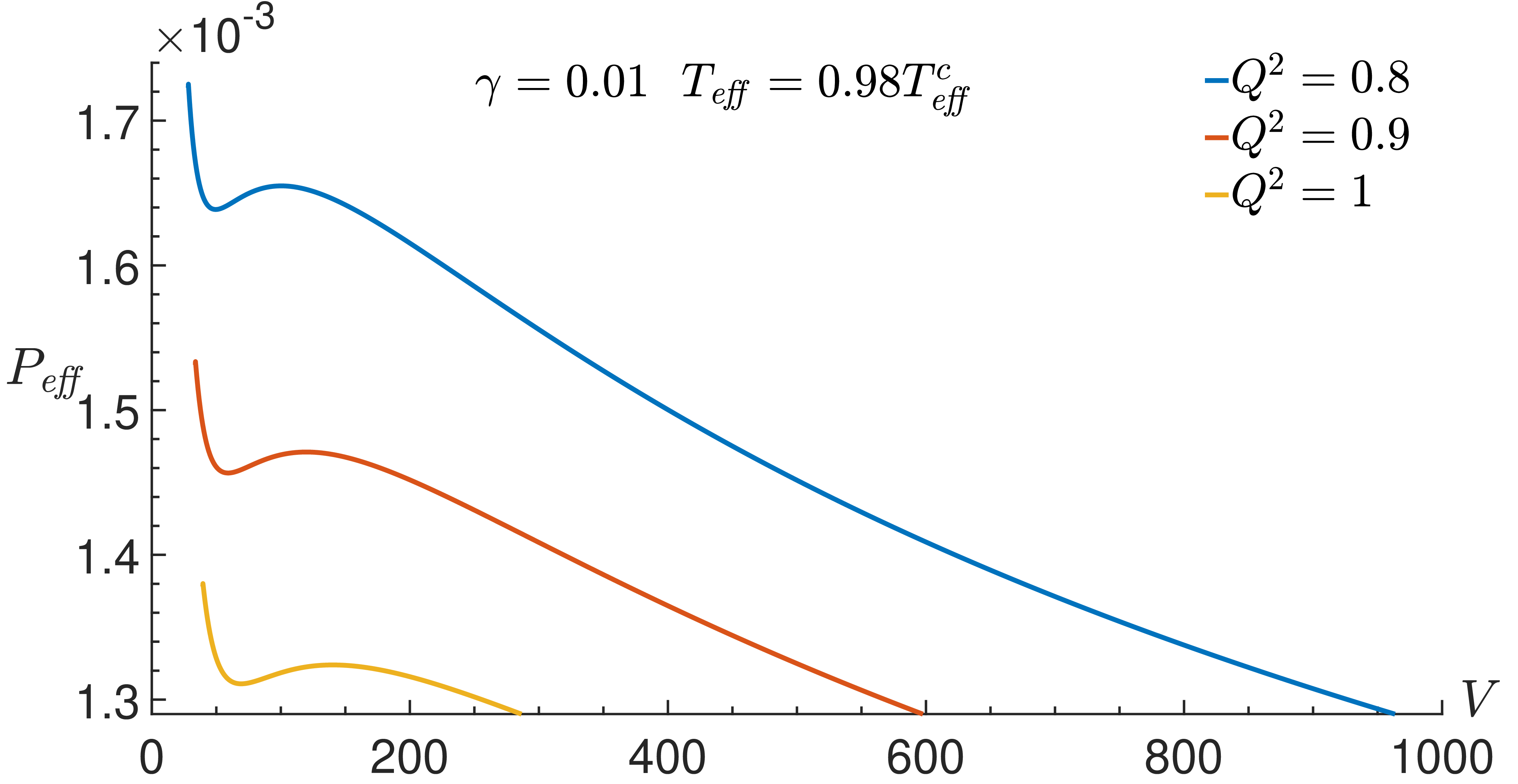}\label{Fig1c}}~
\hspace{0.000001cm}
\subfigure[]{\includegraphics[width=8cm,height=4.5cm]{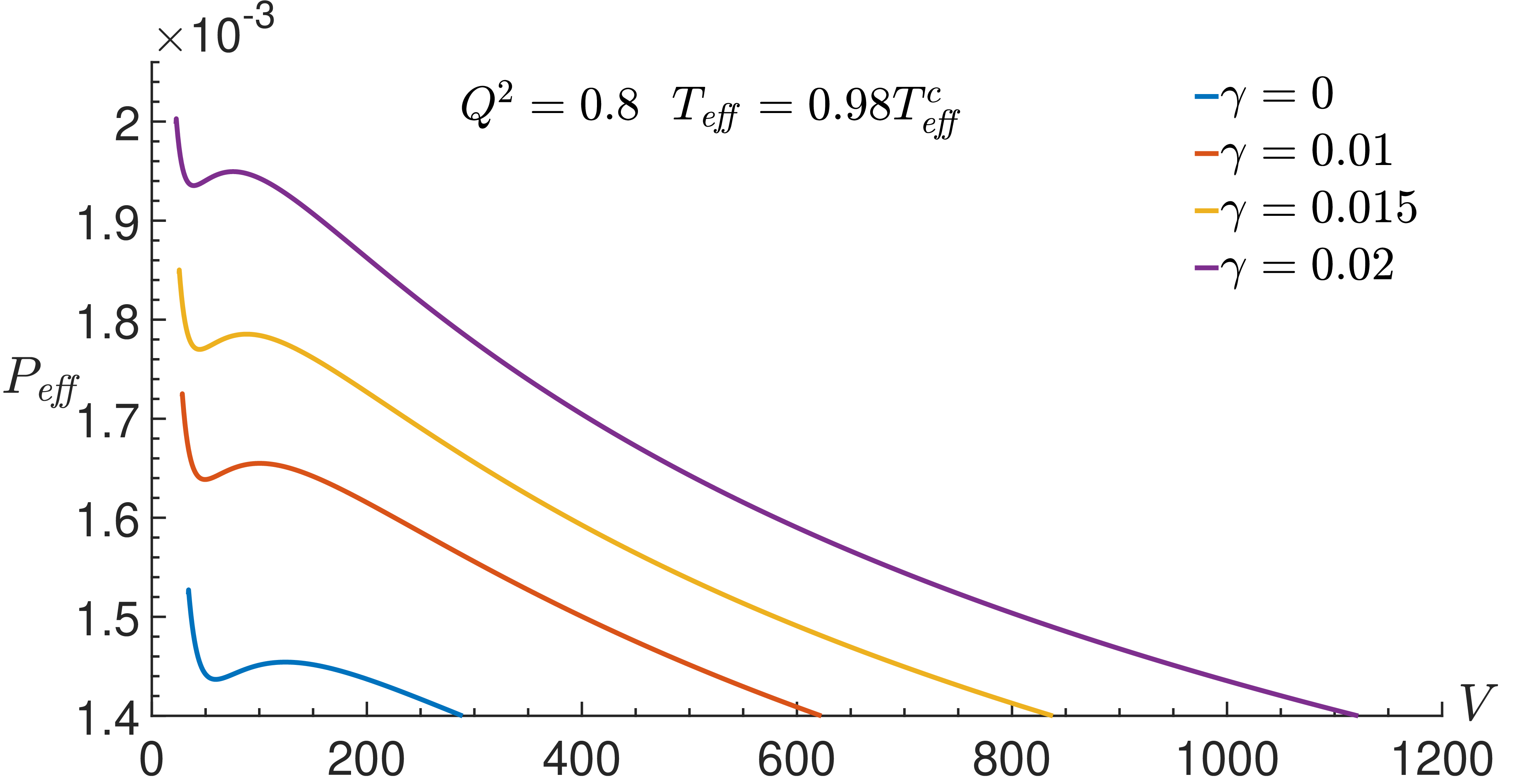}\label{Fig1d}}~
\vskip -1mm \caption{(Color online) Isothermal $P_{eff}-V$ curves with different charges $Q$ and nonlinear parameter $\gamma$.}
\end{figure}

FIG. 3.1 illustrates the influence of charge $Q$ and the nonlinear parameter $\gamma$ on the $P_{eff}-V$ curves. As FIG. 3.1 (b) and (c) shown, under fixed nonlinear parameters $\gamma$, an increase in charge $Q$ causes the isothermal $P_{eff}-V$ curves to shift downward, indicating that this increase reduces the effective pressure $P_{eff}$ of the system. Conversely, as FIG. 3.1 (d) shown, under fixed charge $Q$, an increase in the nonlinear parameter $\gamma$ causes the isothermal $P_{eff}-V$ curves to shift upward, signifying that this increase enhances the effective pressure $P_{eff}$ of the system.

FIG. 3.1 (a) reveals that the isothermal $P_{eff}-V$ curves of the equivalent thermodynamic system exhibit characteristics analogous to those of van der Waals systems or AdS black holes. When the temperature is below the critical temperature (${{T}_{eff}}<T_{eff}^{c}$), a single value of effective pressure corresponds to three distinct volume values along an isotherm. Within this region, segments of the curve with a positive slope (${{\left( \frac{\partial {{P}_{eff}}}{\partial V} \right)}_{{{T}_{eff}}}}>0$) are physically unrealizable as they violate the requirement for thermodynamic equilibrium stability. Therefore, we will determine the stable equilibrium state under fixed temperature ${{T}_{eff}}$ and pressure $P_{eff}$ by applying the minimization condition for the Gibbs free energy.

The critical values of the equivalent thermodynamic system are determined by solving the critical point conditions:
\begin{eqnarray}
{{\left( \frac{\partial {{P}_{eff}}}{\partial V} \right)}_{{{T}_{eff}},Q,\gamma}}={{\left( \frac{{{\partial }^{2}}{{P}_{eff}}}{\partial {{V}^{2}}} \right)}_{{{T}_{eff}},Q,\gamma }}=0 \label{3.1}
\end{eqnarray}
The critical values for varying parameters are provided in TABLE 3.1.

\setcounter{table}{0} 
\renewcommand{\thetable}{3.1} 
\begin{table}[htbp]
\centering
\caption{Critical values of $x_{c}$, $r_{c}$, $T_{eff}^{c}$, $P_{eff}^{c}$, and ${{V}^{c}}$ with fixed charge $Q$ and nonlinear parameter $\gamma$}
\numberwithin{table}{section}
\setlength{\tabcolsep}{12pt} 
\renewcommand{\arraystretch}{1.7} 
\begin{tabular}{c|*{6}{c}} 
\hline
& $\gamma$ & $x_{c}$ & $r_{c}$ & $T_{eff}^{c}$ & $P_{eff}^{c}$ & ${{V}^{c}}$ \\
\hline

\multirow{4}{*}{${{Q}^{2}}=0.8$}
& 0 & 0.77912 & 3.36347 & 0.0206835 & 0.00153323 & 84.0050 \\ \cline{2-7}
& 0.01 & 0.778512 & 3.14254 & 0.0223197 & 0.00174096 & 68.6587 \\ \cline{2-7}
& 0.015 & 0.778115 & 3.01664 & 0.0233457 & 0.00187572 & 60.8158 \\ \cline{2-7}
& 0.02 & 0.777625 & 2.87233 & 0.0245958 & 0.002045 & 52.5870 \\ \hline

\multirow{4}{*}{${{Q}^{2}}=0.9$}
& 0 & 0.77912 & 3.5675 & 0.0195006 & 0.00136287 & 100.2384 \\ \cline{2-7}
& 0.01 & 0.778512 & 3.33316 & 0.0210432 & 0.00154752 & 81.9260 \\ \cline{2-7}
& 0.015 & 0.778115 & 3.19963 & 0.0220105 & 0.00166731 & 72.5680 \\ \cline{2-7}
& 0.02 & 0.777625 & 3.04657 & 0.0231892 & 0.00181778 & 62.7493 \\ \hline

\multirow{4}{*}{${{Q}^{2}}=1$}
& 0 & 0.77912 & 3.76047 & 0.0184999 & 0.00122658 & 117.4002 \\ \cline{2-7}
& 0.01 & 0.778512 & 3.51346 & 0.0199634 & 0.00139277 & 95.9529 \\ \cline{2-7}
& 0.015 & 0.778115 & 3.37269 & 0.0208811 & 0.00150059 & 84.9914 \\ \cline{2-7}
& 0.02 & 0.777625 & 3.21136 & 0.0219992 & 0.001636 & 73.4924 \\ \hline
\end{tabular}
\label{tab:Table. 3.1}
\end{table}

TABLE 3.1 indicates that both the critical pressure $P_{eff}^{c}$ and critical temperature $T_{eff}^{c}$ increase with increasing nonlinear parameter $\gamma$ while holding the charge $Q$ constant. Conversely, for a fixed nonlinear parameter $\gamma$, they decrease with increasing charge $Q$.

The constant-pressure heat capacity of the equivalent thermodynamic system is given by:
\begin{eqnarray}
{{C}_{{{P}_{eff,Q,\gamma }}}}={{T}_{eff}}{{\left( \frac{\partial S}{\partial {{T}_{eff}}} \right)}_{{{P}_{eff}},Q,\gamma }}={{T}_{eff}}\frac{\frac{\partial S}{\partial x}\frac{\partial {{P}_{eff}}}{\partial {{r}_{c}}}-\frac{\partial S}{\partial {{r}_{c}}}\frac{\partial {{P}_{eff}}}{\partial x}}{\frac{\partial {{T}_{eff}}}{\partial x}\frac{\partial {{P}_{eff}}}{\partial {{r}_{c}}}-\frac{\partial {{T}_{eff}}}{\partial {{r}_{c}}}\frac{\partial {{P}_{eff}}}{\partial x}} \label{3.2}
\end{eqnarray}

Using Eq. (3.2), we derive the constant-pressure heat capacity curves (${{C}_{{{P}}}}-{{T}_{eff}}$) at fixed pressure (${{P}_{eff}}=P_{eff}^{c}$) for different charge $Q$ and nonlinear parameter $\gamma$, as shown in FIG 3.2.

\begin{figure}[htbp]
\centering
\subfigure[]{\includegraphics[width=5.9cm,height=3.5cm]{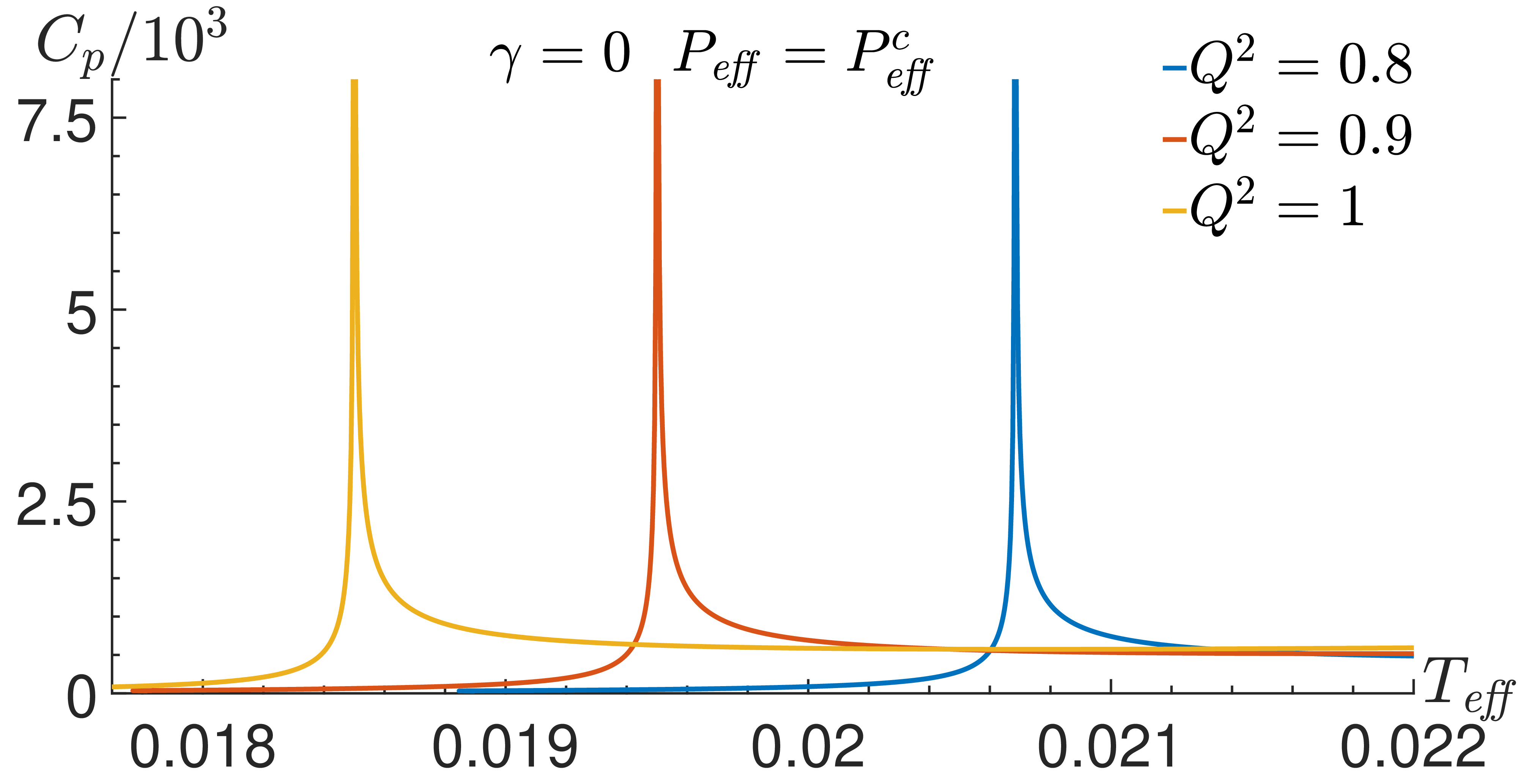}\label{Fig2a}}~
\hspace{0.000001cm}
\subfigure[]{\includegraphics[width=5.9cm,height=3.5cm]{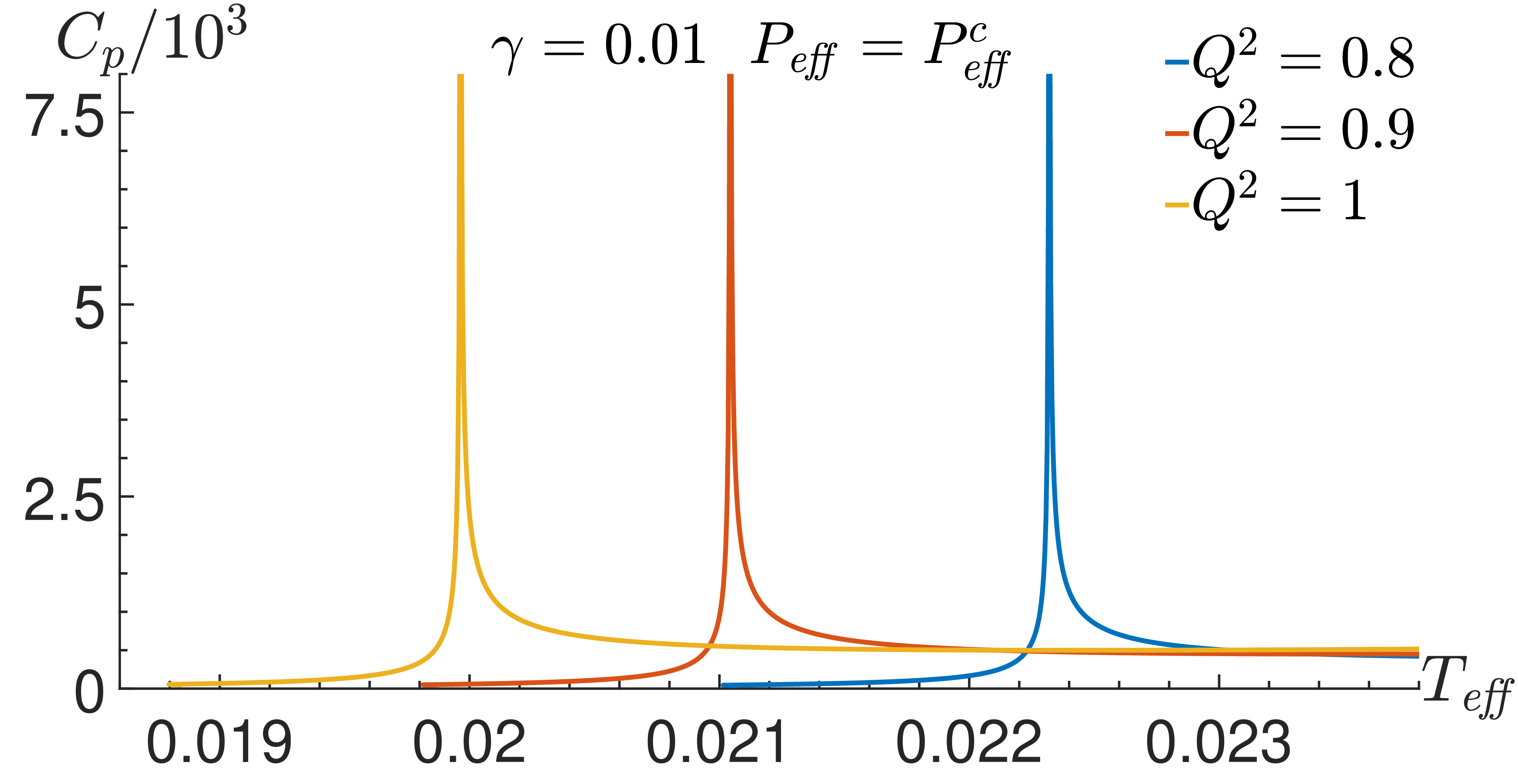}\label{Fig2b}}~
\hspace{0.000001cm}
\subfigure[]{\includegraphics[width=5.9cm,height=3.5cm]{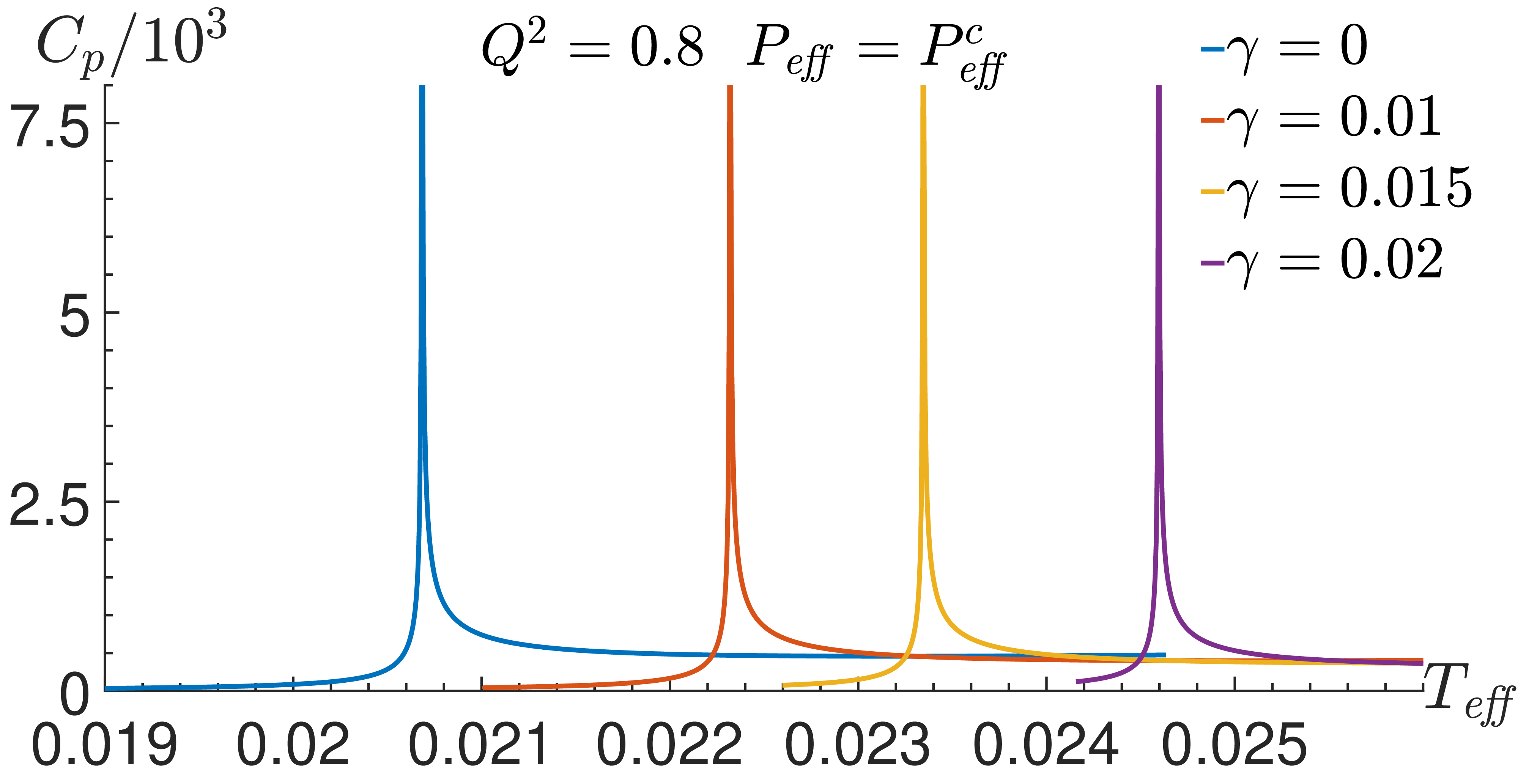}\label{Fig2c}}~
\vskip -1mm \caption{(Color online) ${{C}_{{{P}}}}-{{T}_{eff}}$ curves at ${{P}_{eff}}=P_{eff}^{c}$ for different charge $Q$ and nonlinear parameter $\gamma$.}
\end{figure}

The volume expansion coefficient is defined as:

\begin{eqnarray}
\beta =\frac{1}{V}{{\left( \frac{\partial V}{\partial {{T}_{eff}}} \right)}_{{{P}_{eff}},Q,\gamma }}=\frac{1}{V}\frac{\frac{\partial V}{\partial x}\frac{\partial {{P}_{eff}}}{\partial {{r}_{c}}}-\frac{\partial V}{\partial {{r}_{c}}}\frac{\partial {{P}_{eff}}}{\partial x}}{\frac{\partial {{T}_{eff}}}{\partial x}\frac{\partial {{P}_{eff}}}{\partial {{r}_{c}}}-\frac{\partial {{T}_{eff}}}{\partial {{r}_{c}}}\frac{\partial {{P}_{eff}}}{\partial x}} \label{3.3}
\end{eqnarray}

When ${{P}_{eff}}=P_{eff}^{c}$, the temperature-dependence curves of $\beta -{{T}_{eff}}$ for different charge $Q$ and nonlinear parameter $\gamma$ is shown in FIG 3.3.

\begin{figure}[htbp]
\centering
\subfigure[]{\includegraphics[width=5.9cm,height=3.5cm]{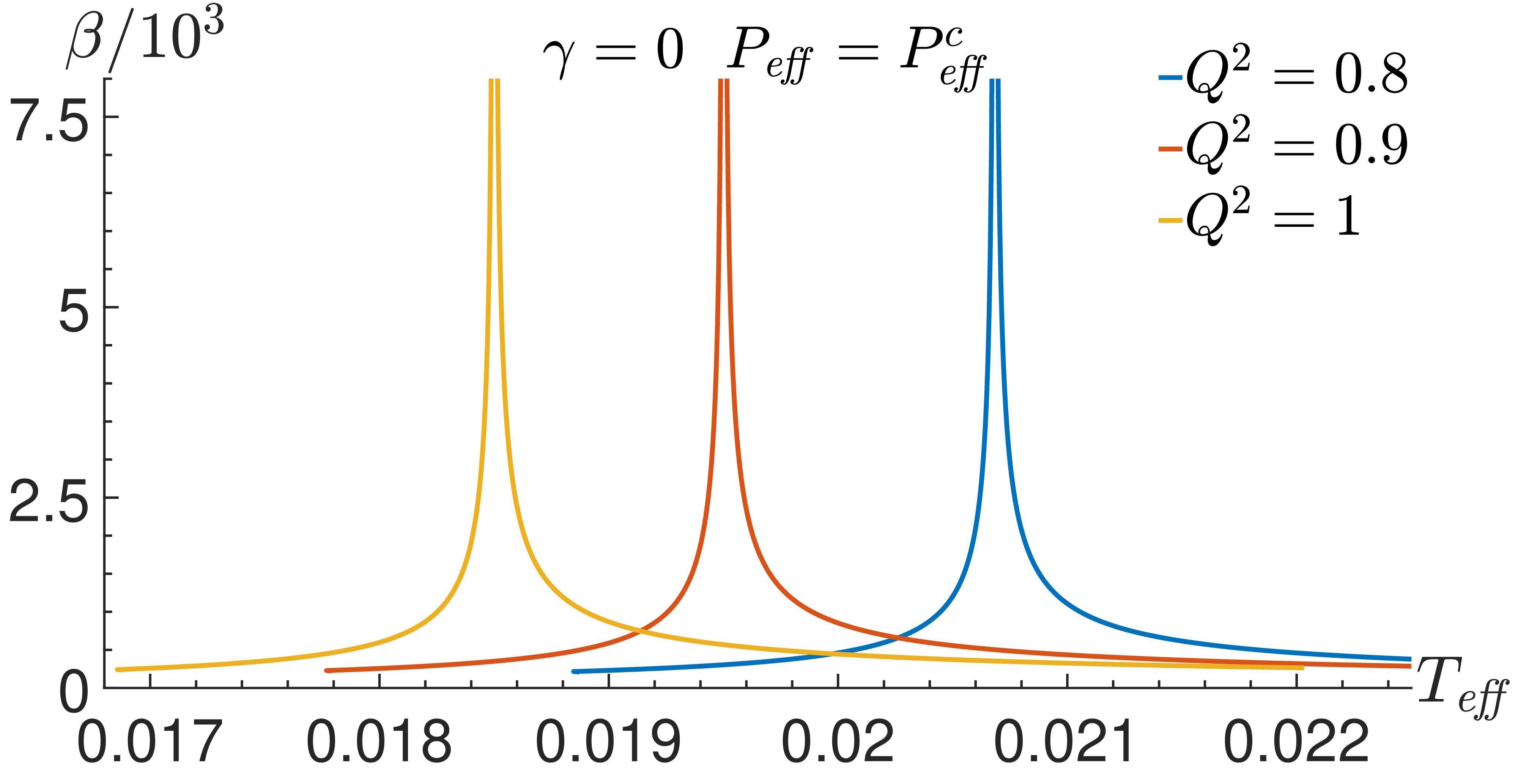}\label{Fig3a}}~
\hspace{0.000001cm}
\subfigure[]{\includegraphics[width=5.9cm,height=3.5cm]{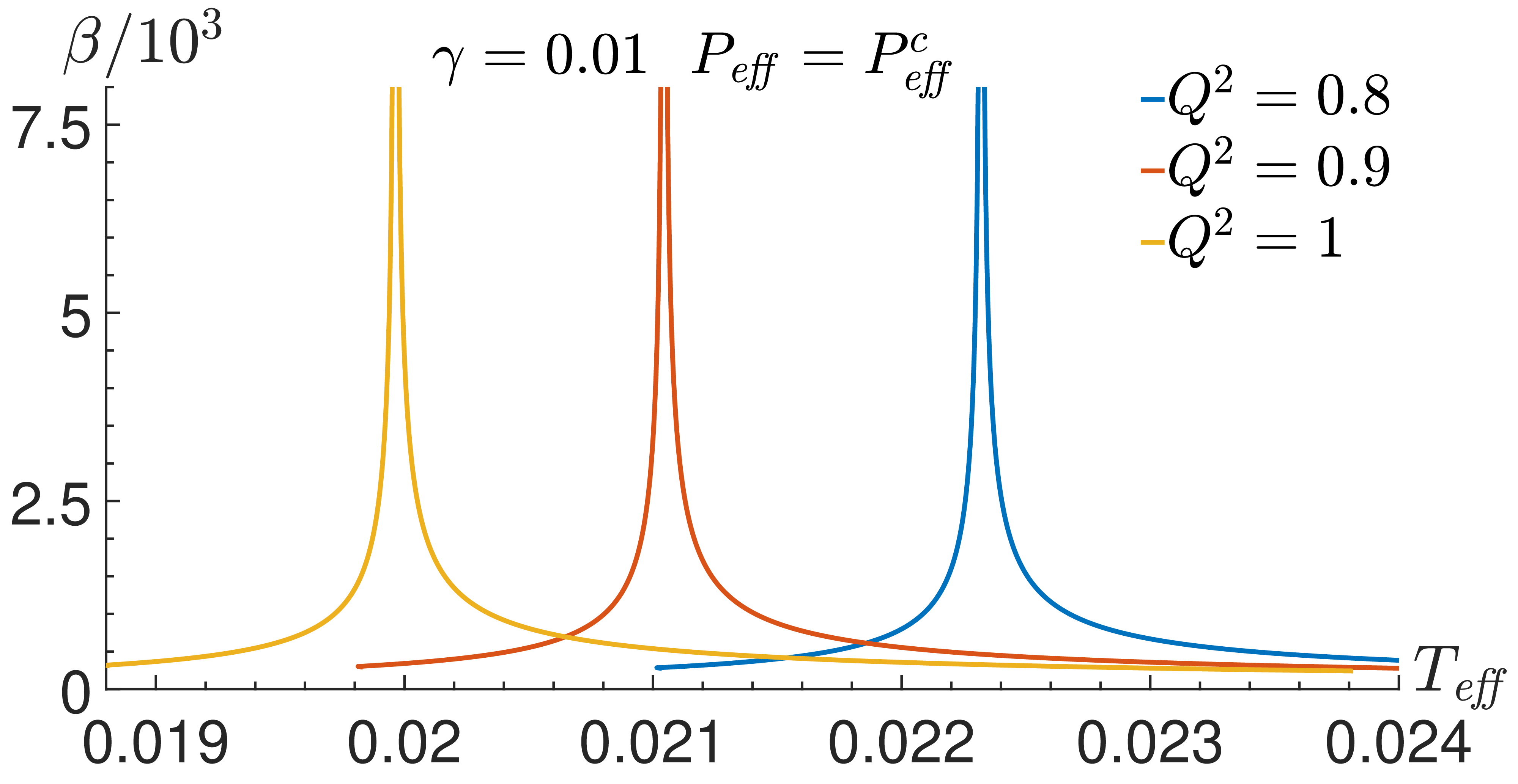}\label{Fig3b}}~
\hspace{0.000001cm}
\subfigure[]{\includegraphics[width=5.9cm,height=3.5cm]{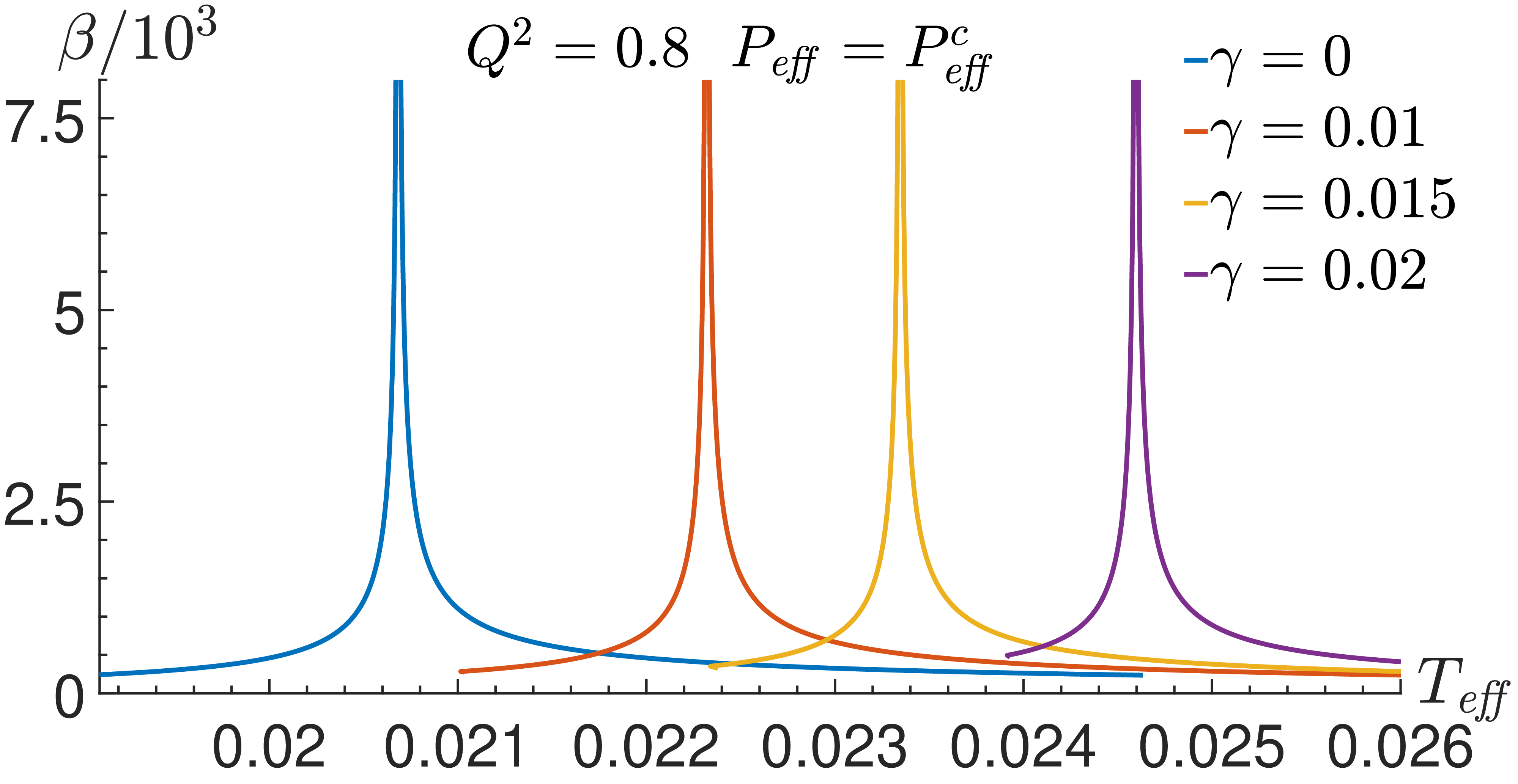}\label{Fig3c}}~
\vskip -1mm \caption{(Color online) $\beta -{{T}_{eff}}$ curves at ${{P}_{eff}}=P_{eff}^{c}$ for different charges $Q$ and nonlinear parameter $\gamma$.}
\end{figure}

The isothermal compressibility is given by:

\begin{eqnarray}
{{\kappa }_{{{T}_{eff}}}}=-\frac{1}{V}{{\left( \frac{\partial V}{\partial {{P}_{eff}}} \right)}_{{{T}_{eff}},Q}}=\frac{1}{V}\frac{\frac{\partial V}{\partial x}\frac{\partial {{T}_{eff}}}{\partial {{r}_{c}}}-\frac{\partial V}{\partial {{r}_{c}}}\frac{\partial {{T}_{eff}}}{\partial x}}{\frac{\partial {{T}_{eff}}}{\partial x}\frac{\partial {{P}_{eff}}}{\partial {{r}_{c}}}-\frac{\partial {{T}_{eff}}}{\partial {{r}_{c}}}\frac{\partial {{P}_{eff}}}{\partial x}} \label{3.4}
\end{eqnarray}

Then we obtain FIG. 3.4, which shows the ${{\kappa}}-{{T}_{eff}}$ curves under different charge $Q$ and nonlinear parameter $\gamma$ at ${{P}_{eff}}=P_{eff}^{c}$.

\begin{figure}[htbp]
\centering
\subfigure[]{\includegraphics[width=5.9cm,height=3.5cm]{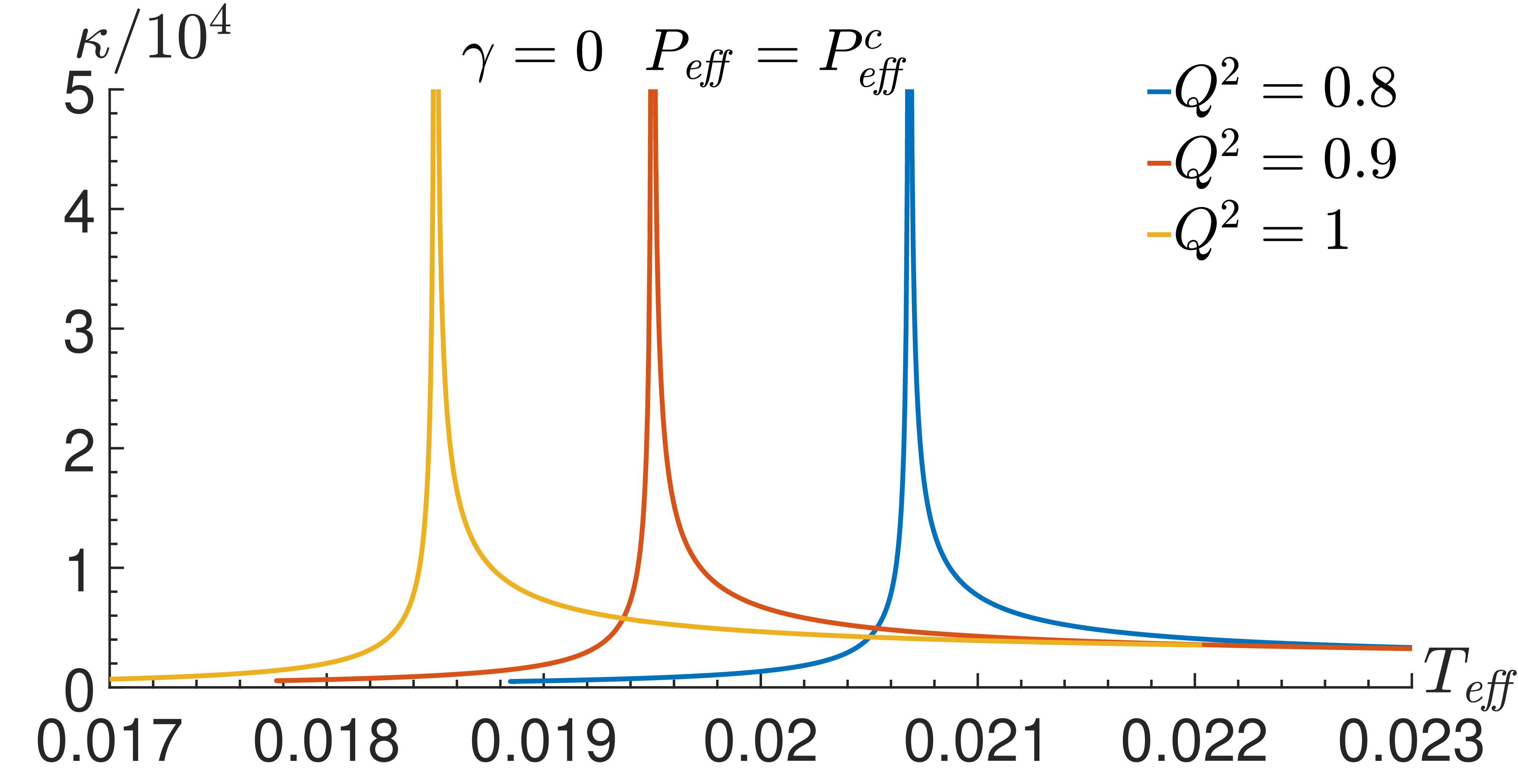}\label{Fig4a}}~
\hspace{0.000001cm}
\subfigure[]{\includegraphics[width=5.9cm,height=3.5cm]{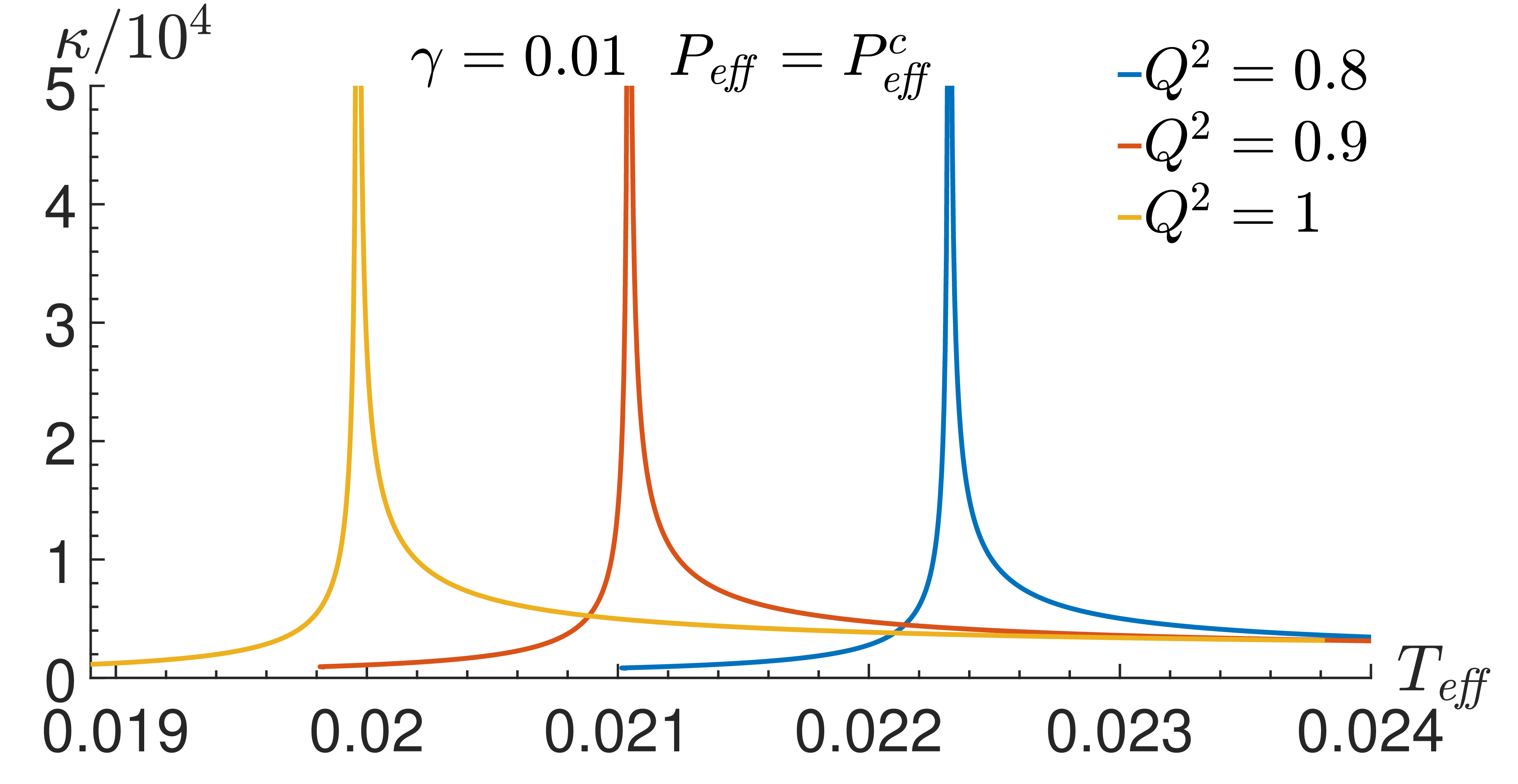}\label{Fig4b}}~
\hspace{0.000001cm}
\subfigure[]{\includegraphics[width=5.9cm,height=3.5cm]{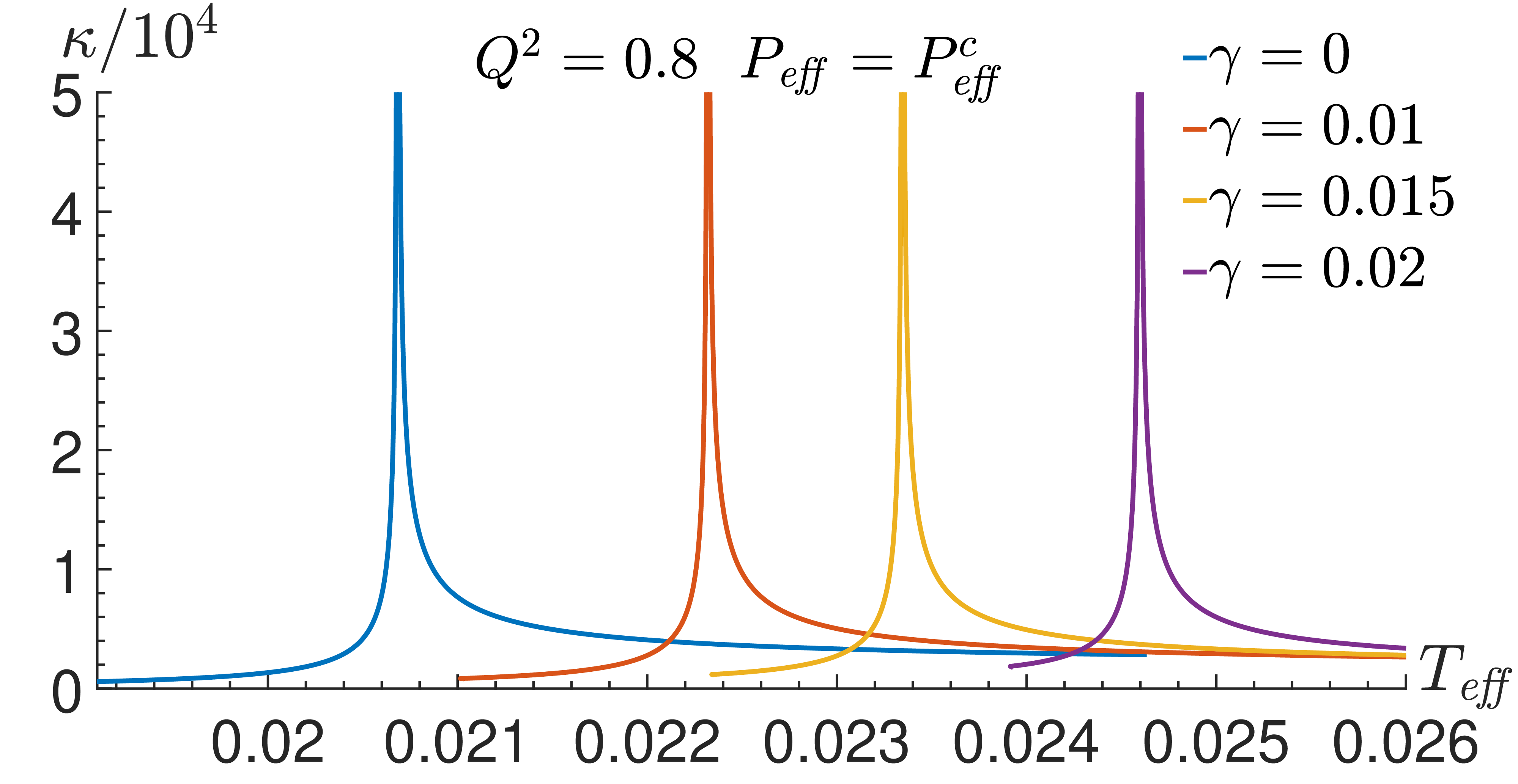}\label{Fig4c}}~
\vskip -1mm \caption{(Color online) ${{\kappa}}-{{T}_{eff}}$ curves at ${{P}_{eff}}=P_{eff}^{c}$ for different charge $Q$ and nonlinear parameter $\gamma$.}
\end{figure}

The total differential of the Gibbs function for the equivalent system is given by:
\begin{eqnarray}
G=M-{{T}_{eff}}S+{{P}_{eff}}V,~~dG=-Sd{{T}_{eff}}+d{{P}_{eff}} \label{3.5}
\end{eqnarray}

\begin{figure}[H]
\centering
\subfigure[]{\includegraphics[width=8cm,height=4.5cm]{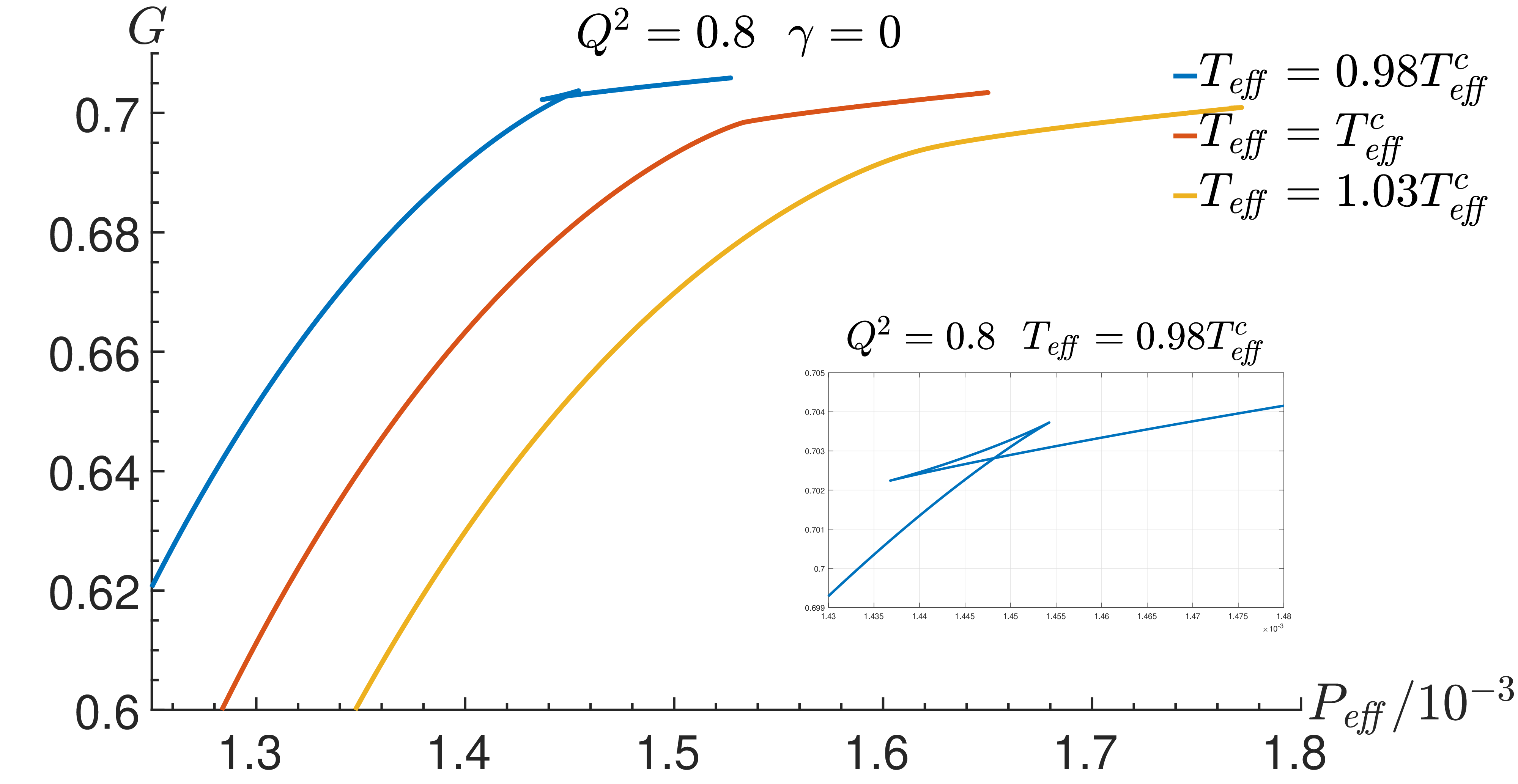}\label{Fig5a}}~
\hspace{0.000001cm}
\subfigure[]{\includegraphics[width=8cm,height=4.5cm]{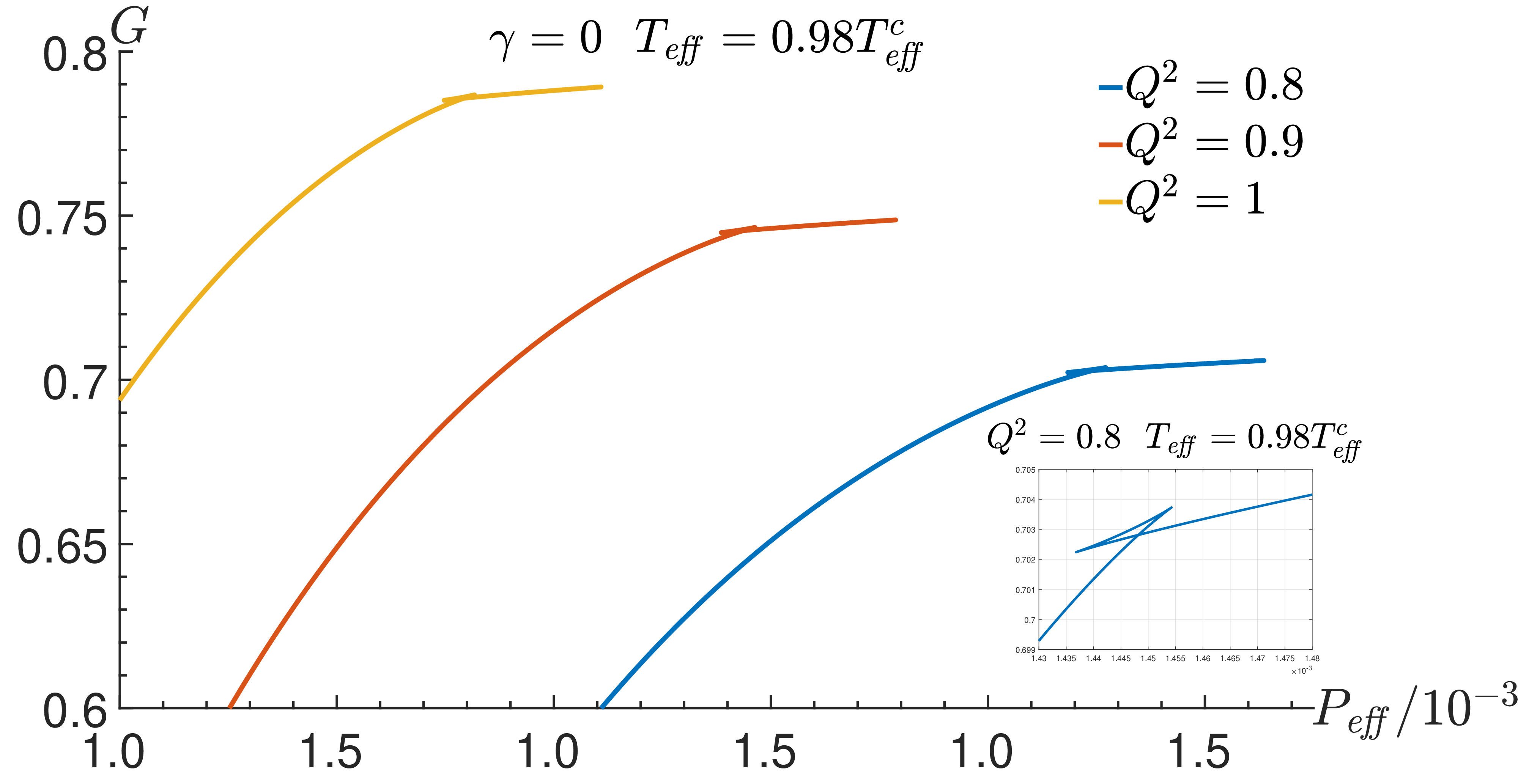}\label{Fig5b}}~
\vskip -1mm \caption{(Color online) Isothermal $G-{{P}_{eff}}$ curves at $\gamma=0$ for different values of charge $Q$.}
\end{figure}

Based on Eq. (3.5), the $G-{{P}_{eff}}$ curves for the equivalent thermodynamic system can be obtained. FIG. 3.5 (a) and 3.6 (b) show that the trends of the $G-{{P}_{eff}}$ curves for nonlinear parameter $\gamma=0$ and $\gamma\neq0$ are similar at ${{T}_{eff}}=T_{eff}^{c}$, exhibiting analogous variation patterns to the van der Waals system. According to Ehrenfest's classification of phase transitions, a transition is classified as second-order if, at the transition point, the Gibbs function and its first derivatives are continuous, while its second derivatives exhibit discontinuities. FIG. 3.1-3.6 indicate that at the equivalent temperature ${{T}_{eff}}=T_{eff}^{c}$, the entropy and thermodynamic volume of spacetime are continuous at the critical point. However, the isobaric heat capacity ${{C}_{{{P}_{eff,Q}}}}$, volume expansion coefficient $\beta$, and isothermal compressibility ${{\kappa }_{{{T}_{eff}}}}$ exhibit discontinuities. This signifies a second-order phase transition in spacetime, with the discontinuity point acting as the second-order transition point. Crucially, this behavior is independent of the value of the nonlinear parameter $\gamma$.

\begin{figure}[htbp]
\centering
\subfigure[]{\includegraphics[width=8cm,height=4.5cm]{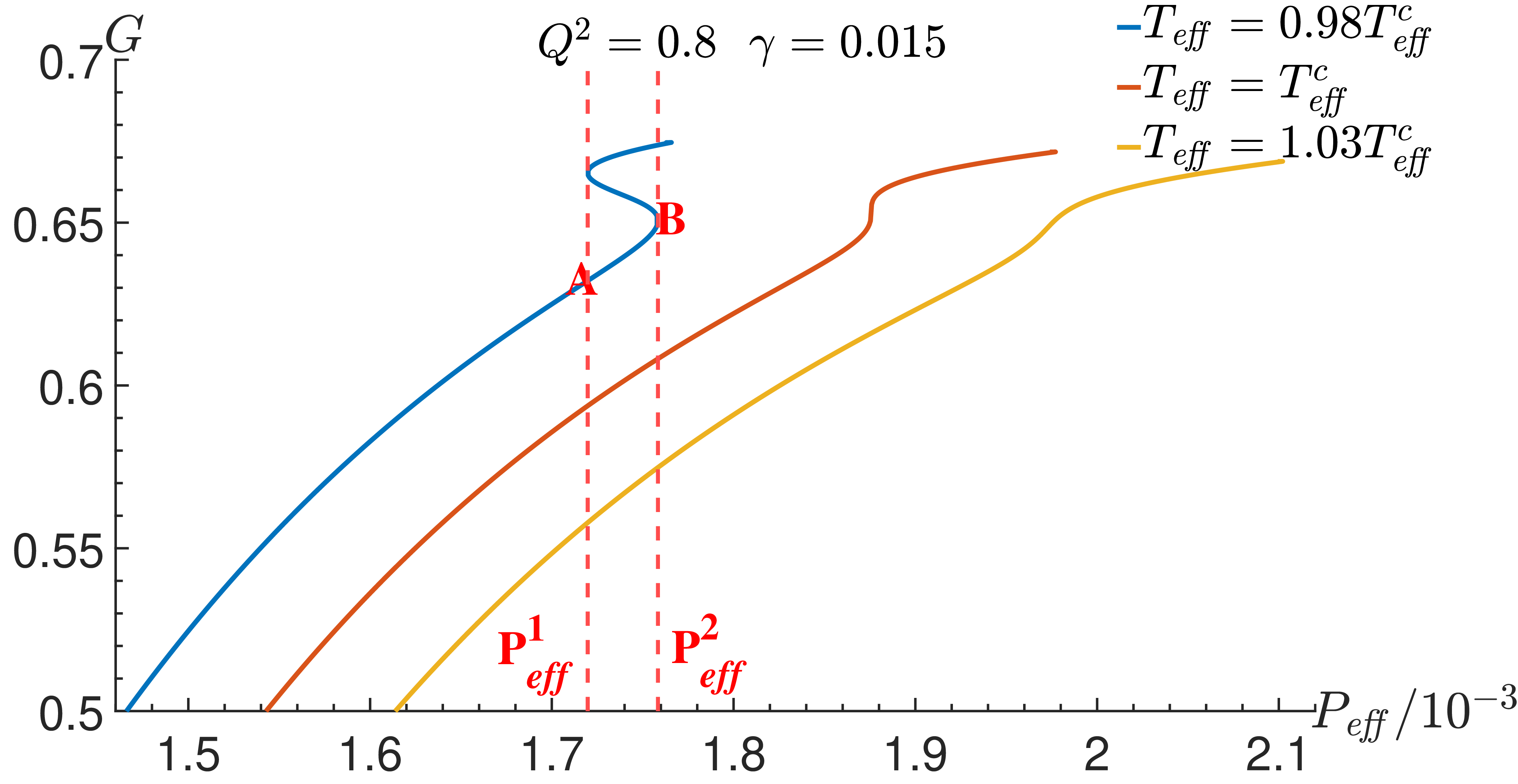}\label{Fig6a}}~
\hspace{0.000001cm}
\subfigure[]{\includegraphics[width=8cm,height=4.5cm]{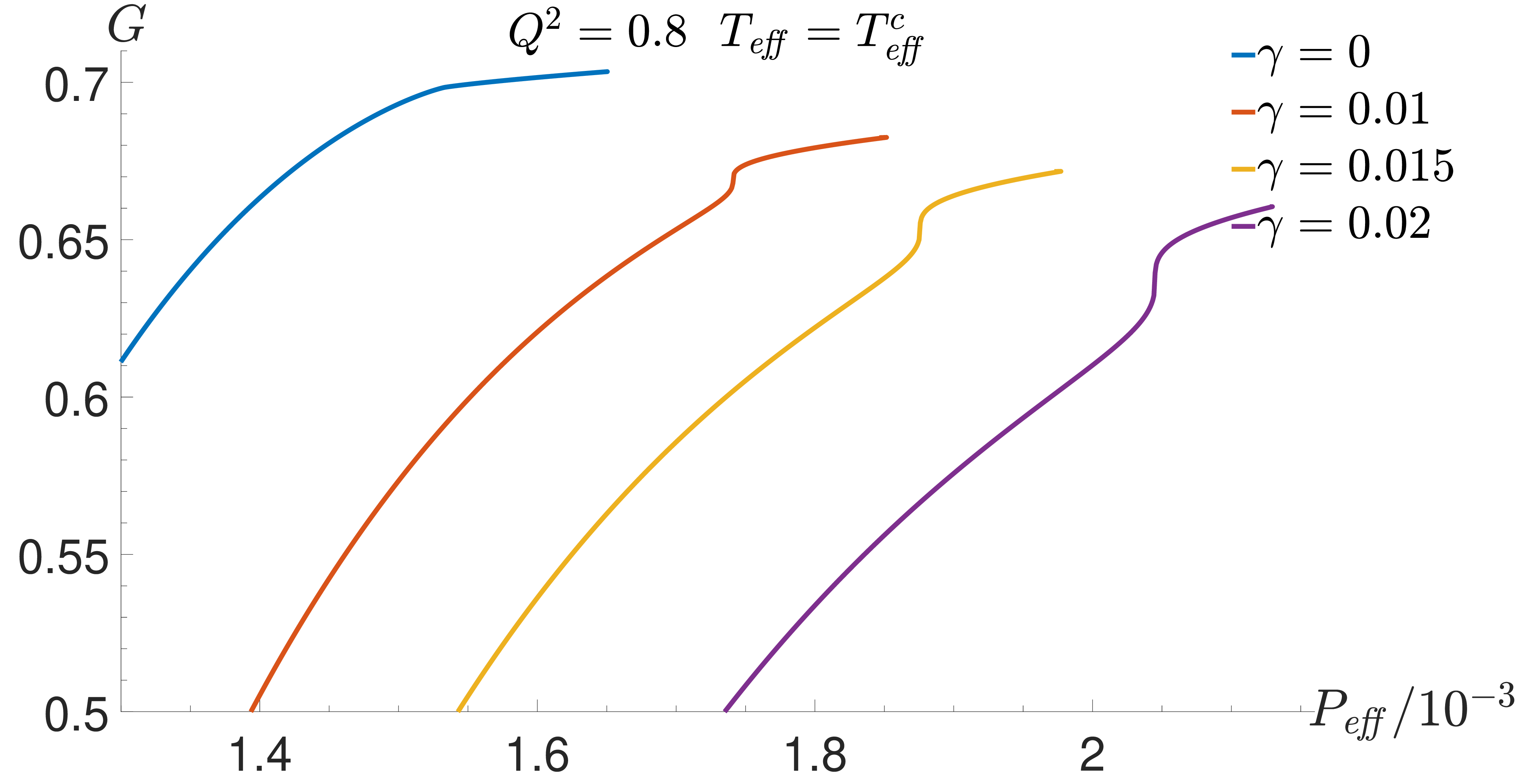}\label{Fig6b}}~
\\
\subfigure[]{\includegraphics[width=8cm,height=4.5cm]{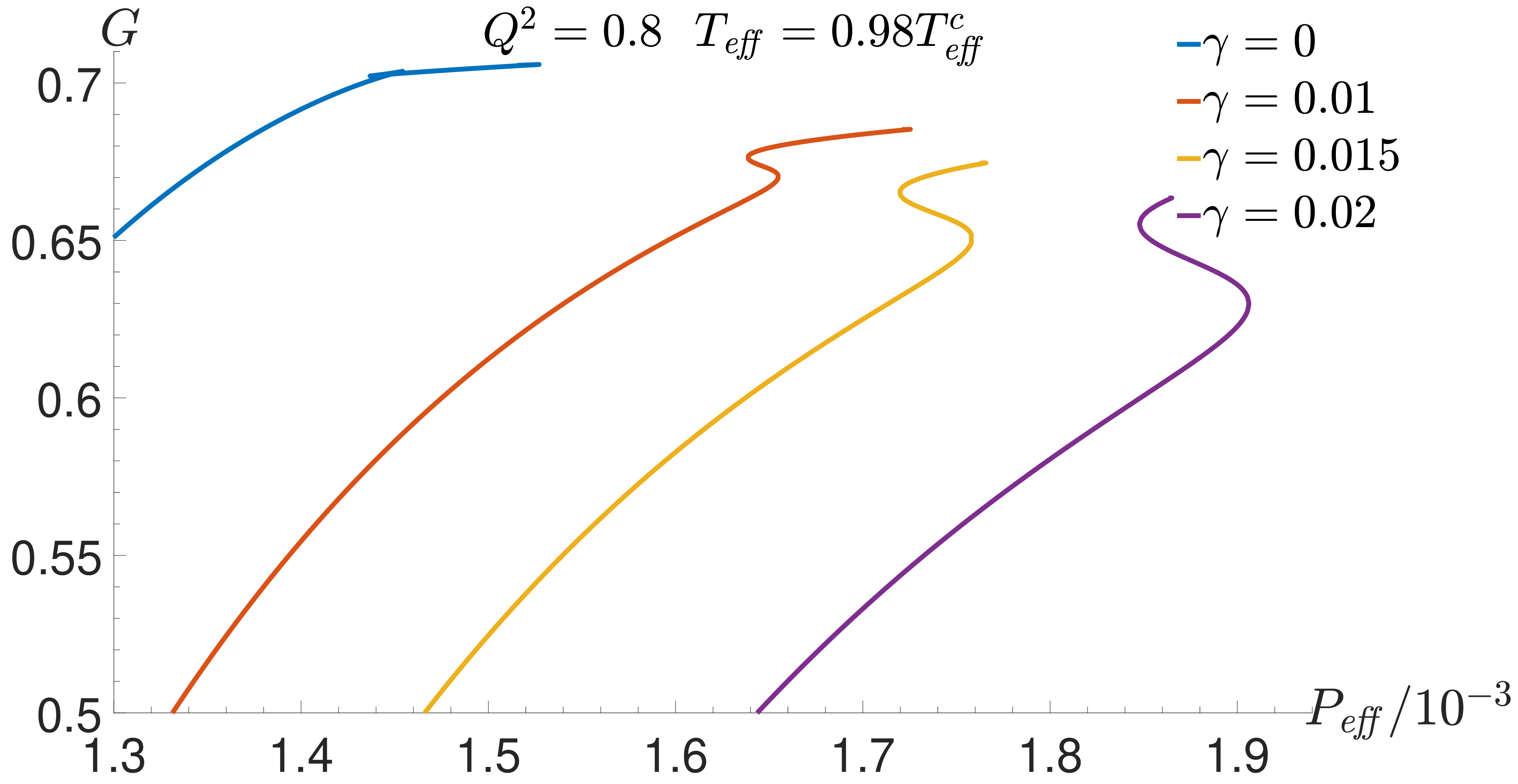}\label{Fig6c}}~
\hspace{0.000001cm}
\subfigure[]{\includegraphics[width=8cm,height=4.5cm]{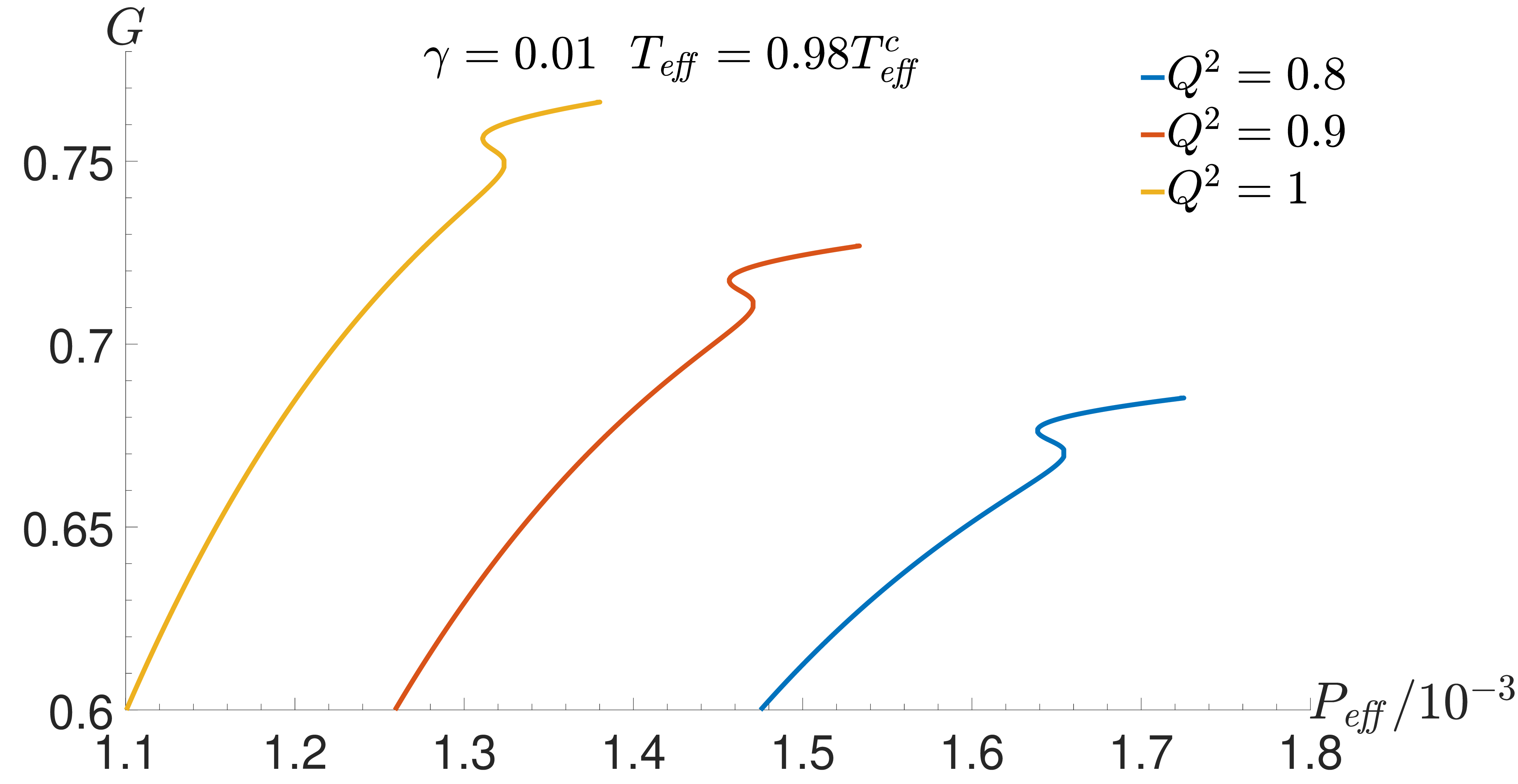}\label{Fig6d}}~
\vskip -1mm \caption{(Color online)Isothermal $G-{{P}_{eff}}$ curves at $\gamma\neq0$ for different values of charge $Q$ and nonlinear parameter $\gamma$.}
\end{figure}

When the equivalent temperature ${{T}_{eff}}<T_{eff}^{c}$, discontinuities occur in the entropy and thermodynamic volume of spacetime. For the nonlinear parameter $\gamma=0$, as seen from the swallowtail $G-{{P}_{eff}}$ curve in FIG. 3.5 (b), the Gibbs function of the system is continuous at the phase transition point, indicating a first-order phase transition in the equivalent system.

FIG. 3.6 shows that for $\gamma\neq0$ and within the pressure range $P_{eff}^{1}<{{P}_{eff}}<P_{eff}^{2}$, a single pressure corresponds to three distinct $G$ values on the isothermal $G-{{P}_{eff}}$ curve. At the specific pressures ${{P}_{eff}}=P_{eff}^{1}$ and ${{P}_{eff}}=P_{eff}^{2}$, a single equivalent pressure corresponds to two different $G$ values. According to the thermodynamic equilibrium stability condition, the Gibbs free energy $G$ must be minimized for a system under constant temperature and pressure. Then under identical temperature and pressure conditions within the interval $P_{eff}^{1}\le {{P}_{eff}}<P_{eff}^{2}$, the equivalent system resides on the AB line. Consequently, when ${{P}_{eff}}=P_{eff}^{2}$, $G$ undergoes a jump, signifying a phase transition in the system. As indicated in FIG. 3.1, under isothermal and isobaric conditions for ${{P}_{eff}}<P_{eff}^{c}$, both the entropy and volume are discontinuous. Applying Ehrenfest's classification of phase transitions, this implies that for $\gamma\neq0$, the phase transition occurring at ${{P}_{eff}}=P_{eff}^{2}$ in the equivalent system is a zeroth-order phase transition.

\section{Topology of the equivalent thermodynamic system}\label{four}

The Helmholtz free energy of the equivalent system is given by:

\begin{eqnarray}
F({{r}_{c}},x)=M-{{T}_{eff}}S \label{4.1}
\end{eqnarray}

In order to construct the thermodynamical topology, the vector field mapping $\phi$: $X=\left. (s,\theta ) \right|0<s<\infty $, $0<\theta <\pi \to {{\mathbb{R}}^{2}}$ is defined as follows \cite{50,51,52,53,54,55}:

\begin{eqnarray}
\phi (S,\theta )=\left( \frac{\partial F}{\partial S},-\cot \theta \csc \theta  \right) \label{4.2}
\end{eqnarray}

When the parameter $S$ is used to characterize the AdS black hole, it becomes the first parameter in the domain of the mapping $\phi$. The parameter $\theta$ serves as an auxiliary function and is utilized to construct the second component of the mapping. The component ${{\phi }^{\theta }}$ diverges at $\theta=0,\pi$, resulting in an outward-pointing vector field orientation at these points. It is readily apparent that the zero point of $\phi$ corresponds to black hole with temperature ${{T}_{eff}}={{\tau }^{-1}}$ as $\theta =\pi /2$. Consequently, the zero points of the mapping can be employed to characterize black hole solutions for a given parameter $\tau$.

Based on Duan's $\phi$-mapping topological current theory \cite{21}, the zero points of the mapping $\phi$ are associated with a topological number. This topological number is calculated as the weighted sum of the zero points of $\phi$. The weight assigned to each zero point, known as its topological charge, is determined by its nature: saddle points carry a weight of $-1$, while extremal points carry a weight of $+1$. The topological current can be expressed in the following form:

\begin{eqnarray}
{{j}^{\mu }}=\frac{1}{2\pi }{{e}^{\mu \nu \rho }}{{\varepsilon }_{ab}}{{\partial }_{\nu }}{{n}^{a}}{{\partial }_{\rho }}{{n}^{b}},~~\mu ,\nu ,\rho =0,1,2,~~a,b=1,2 \label{4.3}
\end{eqnarray}

The unit vector $n$ is defined as: $n=({{n}^{s}},{{n}^{\theta }})$, where ${{n}^{s}}={{\phi }^{s}}/\left\| \phi  \right\|$ and ${{n}^{\theta }}={{\phi }^{\theta }}/\left\| \phi  \right\|$. It follows directly that the topological current ${{j}^{\mu }}$ is identically conserved: ${{\partial }_{\mu }}{{j}^{\mu }}=0$. Consequently, the topological current ${{j}^{\mu }}$ is proportional to a delta function in the field configuration space:

\begin{eqnarray}
{{j}^{\mu }}={{\delta }^{2}}(\overset{\scriptscriptstyle\rightharpoonup}{\phi }){{J}^{\mu }}\left( \frac{\phi }{x} \right) \label{4.4}
\end{eqnarray}

where the 3-dimensional Jacobian ${{J}^{\mu }}\left( \frac{\phi }{x} \right)$ is defined as: ${{\varepsilon }^{ab}}{{J}^{\mu }}\left( \frac{\phi }{x} \right)={{\varepsilon }^{\mu \nu \rho }}{{\partial }_{\nu }}{{\phi }^{a}}{{\partial }_{\rho }}{{\phi }^{b}}$. It is straightforward to show that ${{j}^{\mu }}$ vanishes only if ${{\phi }^{a}}({{x}_{i}})=0$. Consequently, the topological number $W$ within a parameter region $\Sigma$ is given by:

\begin{eqnarray}
W=\int_{\Sigma }{{{j}^{0}}}{{d}^{2}}x=\sum\limits_{i=1}^{{\bar{N}}}{{{\beta }_{i}}}{{\eta }_{i}}=\sum\limits_{i=1}^{{\bar{N}}}{{{w}_{i}}} \label{4.5}
\end{eqnarray}

Here, ${{j}^{0}}=\sum\nolimits_{i=1}^{{\bar{N}}}{{{\beta }_{i}}{{\eta }_{i}}{{\delta }^{2}}(\vec{x}-{{{\vec{z}}}_{i}})}$ represents the topological current density. ${{\beta }_{i}}$ denotes the Hopf index, which counts the number of the loops that ${{\phi }^{a}}$ makes in the vector $\phi$ space when ${{x}^{\mu }}$ goes around the zero point ${{z}_{i}}$. Crucially, this index is always a positive integer. ${{\eta }_{i}}=sign{{J}^{0}}{{(\phi /x)}_{{{z}_{i}}}}=\pm 1$ is the Brouwer degree. The ${{w}_{i}}$ is the winding number associated with the $i$-th zero point within a given region and its value is topologically invariant, independent of the specific shape of the region. Research has established that this winding number is determined by the (un)stable black hole solutions.

Based on Eq. (4.5), when the charge $Q$ and the nonlinear parameter $\gamma$ of the system are held constant, the following relations hold:

\begin{eqnarray}
{{\phi }^{S}}&=&\frac{\partial F}{\partial S} \notag \\
&=&\frac{1-x}{4\pi {{r}_{c}}{{x}^{5}}}\left\{ \left[ (1+x)(1+{{x}^{3}})-2{{x}^{2}} \right] \right.-\frac{{{Q}^{2}}}{r_{c}^{2}{{x}^{2}}}\left[ (1+x+{{x}^{2}})(1+{{x}^{4}})-2{{x}^{3}} \right]+\left. \frac{{{Q}^{2}}\gamma (5-3{{x}^{3}}+3{{x}^{8}}-5{{x}^{11}})}{5r_{c}^{2}{{x}^{11}}(1-x)} \right\}-\frac{1}{\tau } \notag \\
{{\phi }^{\theta }}&=&-cot\theta csc\theta \label{4.6}
\end{eqnarray}

Zero points are determined by ${{\phi }^{S}}=0$. We solve it and get:

\begin{eqnarray}
\frac{1}{\tau }=\frac{1-x}{4\pi {{r}_{c}}{{x}^{5}}}\left\{ \left[ (1+x)(1+{{x}^{3}})-2{{x}^{2}} \right] \right.-\frac{{{Q}^{2}}}{r_{c}^{2}{{x}^{2}}}\left. \left[ (1+x+{{x}^{2}})(1+{{x}^{4}})-2{{x}^{3}} \right]+\frac{{{Q}^{2}}\gamma (5-3{{x}^{3}}+3{{x}^{8}}-5{{x}^{11}})}{5r_{c}^{2}{{x}^{11}}(1-x)} \right\} \label{4.7}
\end{eqnarray}

\begin{figure}[http]
\centering
\subfigure[]{\includegraphics[width=8cm,height=4.5cm]{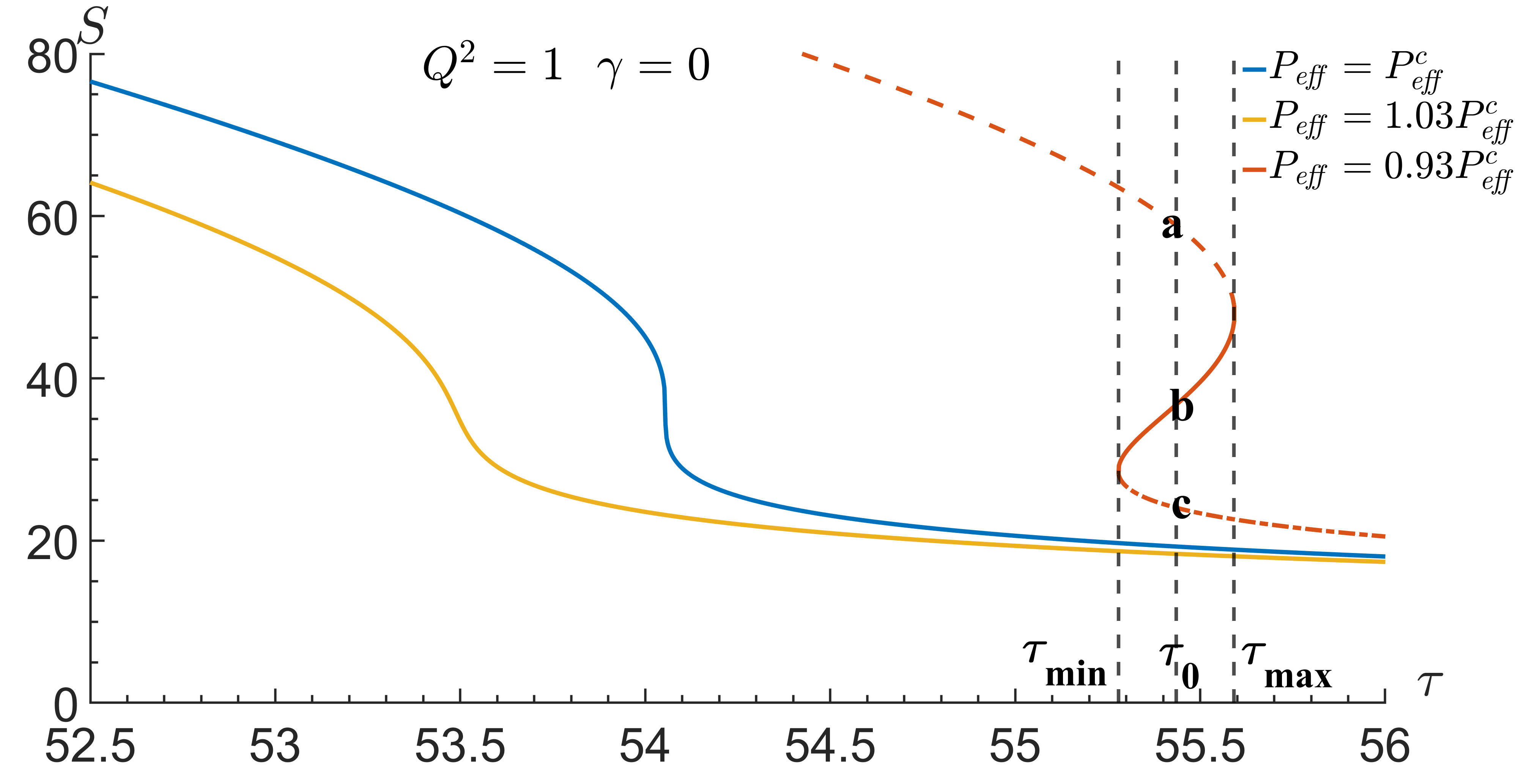}\label{Fig7a}}~
\hspace{0.000001cm}
\subfigure[]{\includegraphics[width=8cm,height=4.5cm]{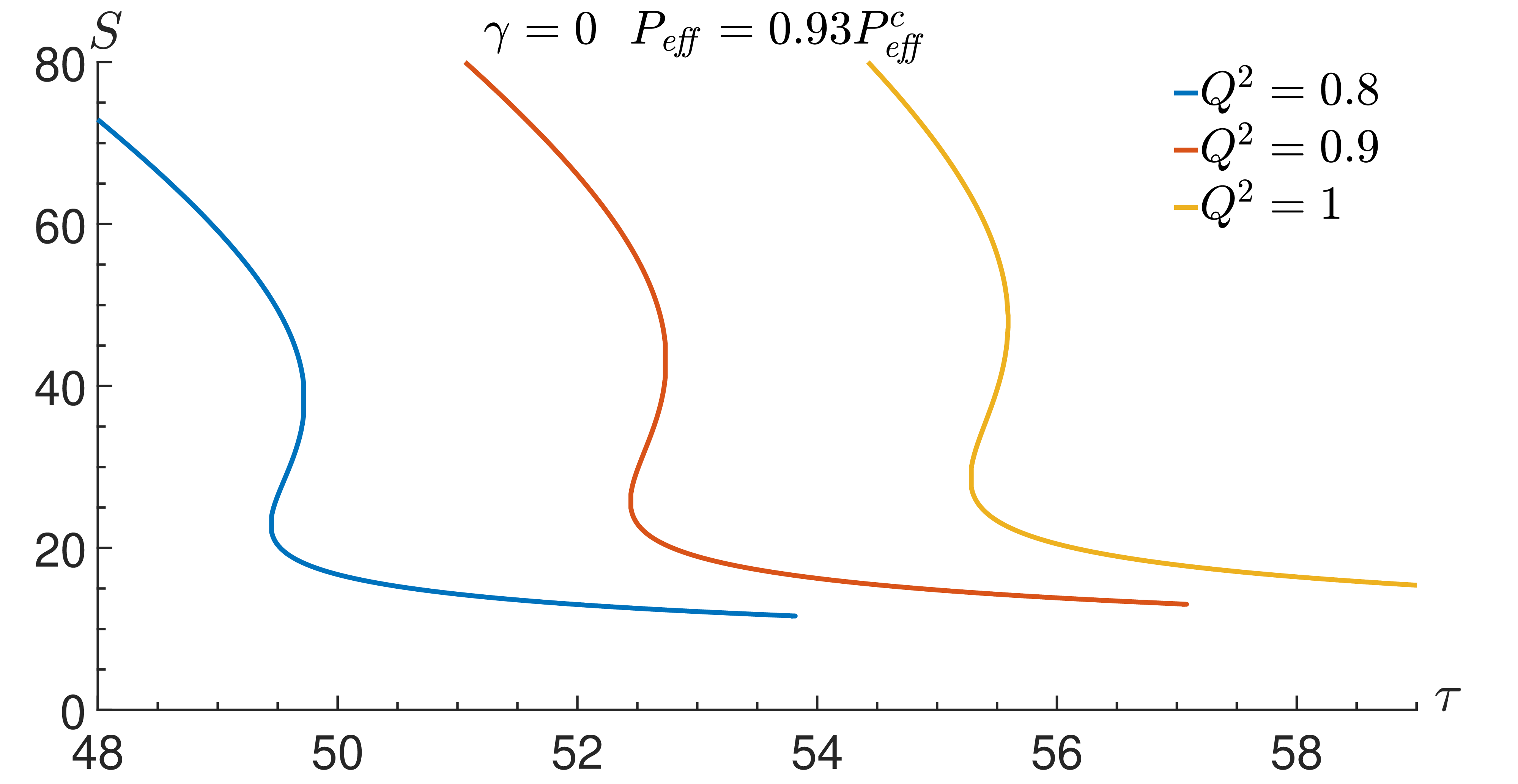}\label{Fig7b}}~
\\
\subfigure[]{\includegraphics[width=8cm,height=4.5cm]{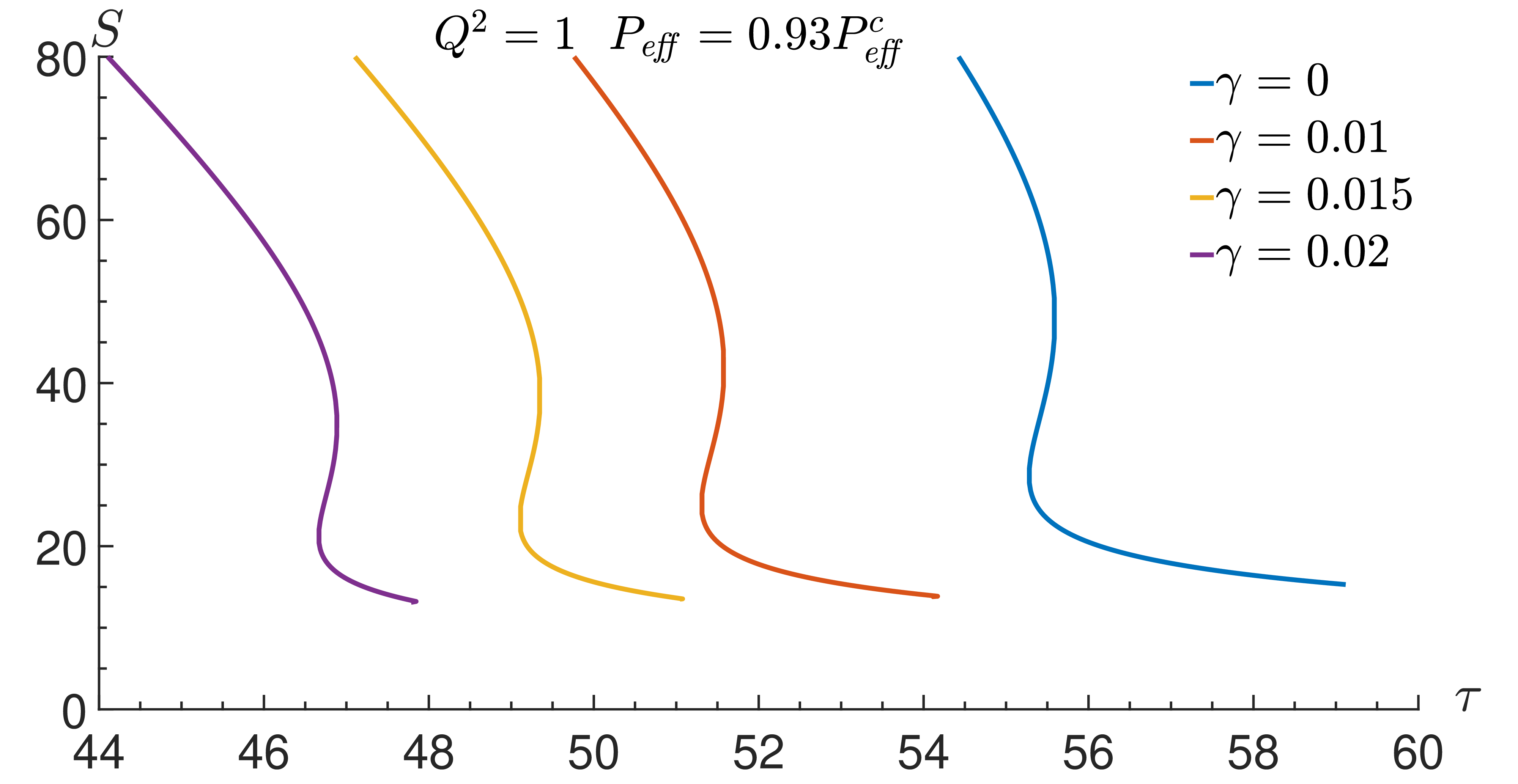}\label{Fig7c}}~
\hspace{0.000001cm}
\subfigure[]{\includegraphics[width=8cm,height=4.5cm]{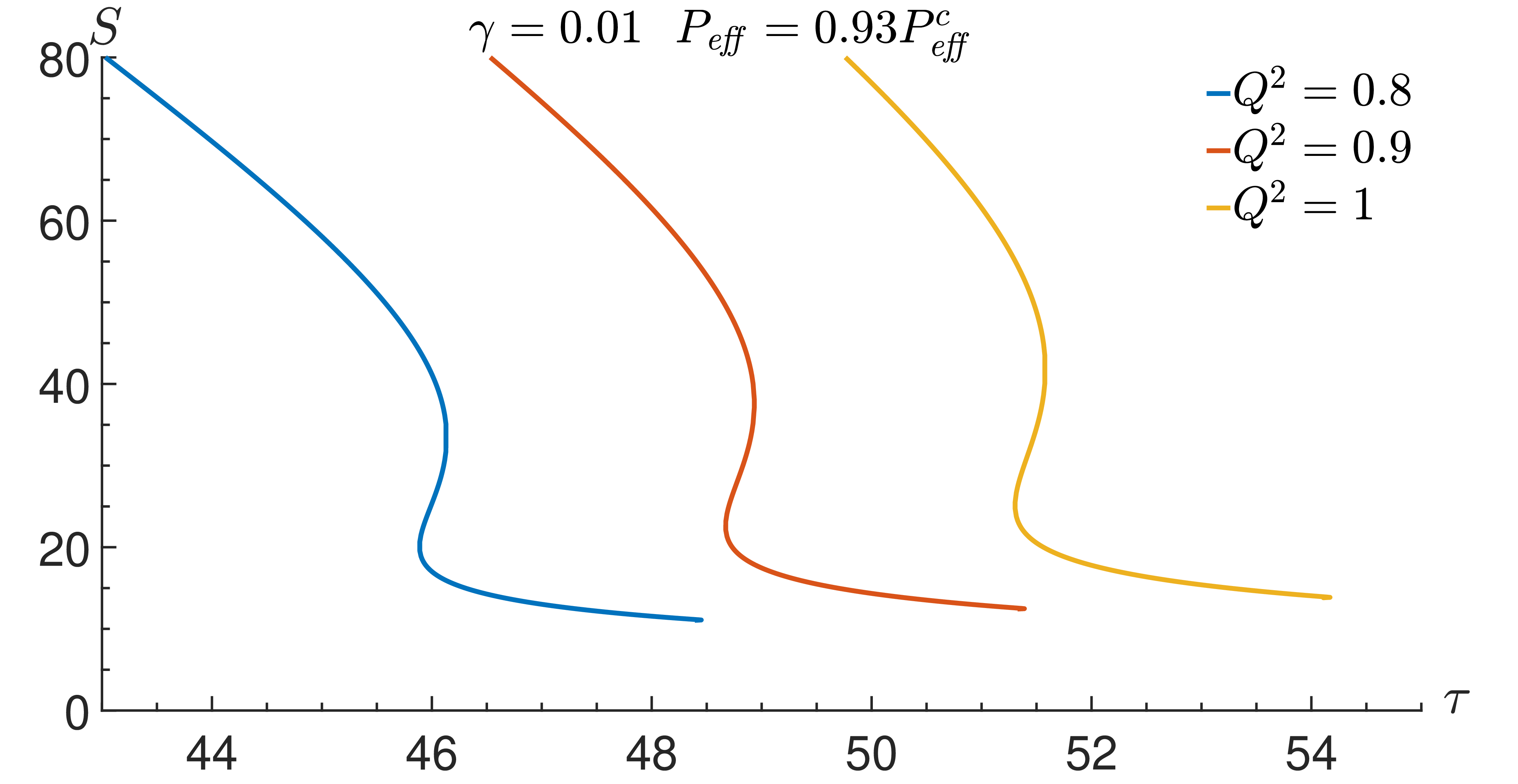}\label{Fig7d}}~
\vskip -1mm \caption{(Color online)Zero points of ${\phi }^{S}$ in the diagram of $S-\tau$ for different values of charge $Q$ and nonlinear parameter $\gamma$.}
\end{figure}
It is obvious that there are two extremes ${{\tau }_{\min }}$ and ${{\tau }_{\max }}$. As $Q$=1, ${{P}_{eff}}=P_{eff}^{c}$, ${{\tau }_{\min }}={{\tau }_{\max }}={{\tau }_{c}}=1$. Here, ${{\tau }_{\min }}$ corresponds to the generation point, and ${{\tau }_{\max }}$ corresponds to the annihilation point. Note that the generation point satisfy the constraint conditions: $\frac{\partial S}{\partial \tau }=0$, $\frac{{{\partial }^{2}}S}{\partial {{\tau }^{2}}}>0$, while the annihilation point obeys: $\frac{\partial S}{\partial \tau }=0$, $\frac{{{\partial }^{2}}S}{\partial {{\tau }^{2}}}<0$. The zero points of ${{\phi }^{S}}$ in the $S-\tau$ diagram for different charge $Q$ and nonlinear parameter $\gamma$ are displayed in FIG. 4.1.

As shown in FIG. 4.1, for ${{P}_{eff}}<P_{eff}^{c}$ and within the temperature range ${{\tau }_{\min }}<\tau <{{\tau }_{\max }}$, there are three intersection points within the coexistence region of two horizons for the dS spacetime in the canonical ensemble. The dashed, solid, and dash-dotted curves correspond to the low-potential small black hole (LPSB), intermediate-potential black hole (IPBH), and high-potential large black hole (HPBH), respectively. These represent stable, unstable, and stable black hole states. The corresponding winding numbers are +1,-1,+1, yielding a topological number $W=+1-1+1=+1$ for ${{\tau }_{\min }}<\tau <{{\tau }_{\max }}$. The intersection points satisfy the condition $\tau =1/{{T}_{eff}}$ precisely at the boundaries $\tau ={{\tau }_{\min }},{{\tau }_{\max }}$. For temperatures outside this range ($\tau <{{\tau }_{\min }}$ or $\tau >{{\tau }_{\max }}$), these points disappear, and the three distinct black hole branches coalesce into a single stable black hole branch. Consequently, the topological number remains $W=+1$.

When ${{\tau }_{\min }}<{{\tau }_{0}}<{{\tau }_{\max }}$, the vertical line $\tau=\tau_{0}$ intersects the $S-\tau$ curve at points a, b, and c. At points a and c, the slope of the $S-\tau$ curve satisfies $\frac{\partial S}{\partial \tau }<0$. These correspond to thermodynamically stable states, and its topological charge $\omega=1$. At point b, the slope satisfies $\frac{\partial S}{\partial \tau }>0$, indicating a thermodynamically unstable state with topological charge $\omega=-1$. The total topological number for the system is therefore $W=+1-1+1=+1$. This configuration satisfies the equilibrium stability requirements for the equivalent thermodynamic system. And for $S-\tau$ curves at pressure ${{P}_{eff}}\ge P_{eff}^{c}$ where $\frac{\partial S}{\partial \tau }<0$, the equivalent thermodynamic system resides in a single stable state. Consequently, the topological number also is $W=+1$. That is the topological number of the equivalent thermodynamic system always is $W=+1$ for different charge $Q$ and nonlinear parameter $\gamma$ at full equivalent pressure range. These findings are consistent with the results reported in Refs \cite{27,55,56}.

Based on Eq. (4.6), we can plot the vector field diagrams on a segment of the ($S-\theta$) plane in Euler-Heisenberg dS spacetime for different values of equivalent pressure, with the charge $Q=1$ and fixed nonlinear parameter $\gamma=0,0.01$, as shown in FIG. 4.2.
\begin{figure}[http]
\centering
\subfigure[]{\includegraphics[width=8cm,height=4.5cm]{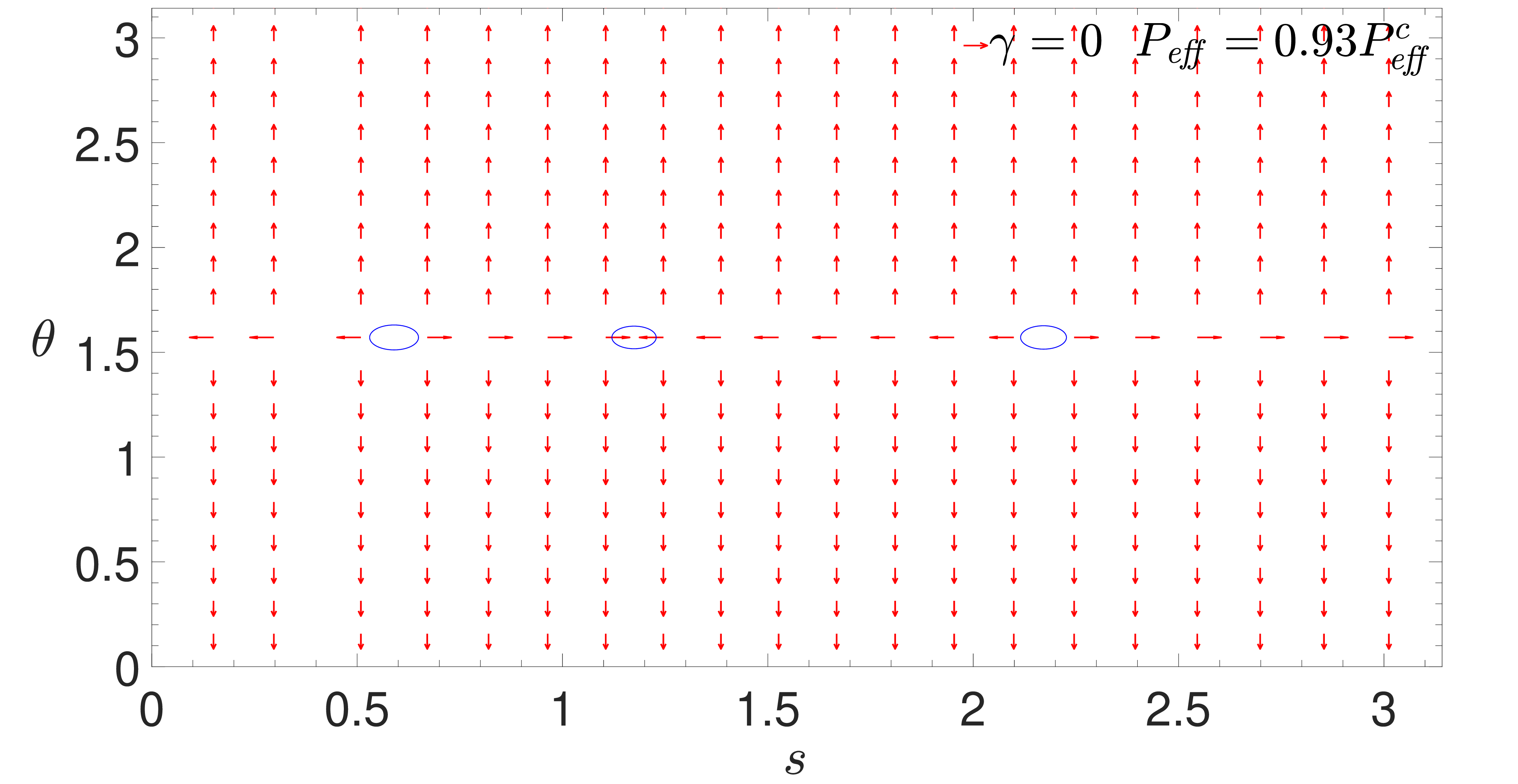}\label{Fig8a}}~
\hspace{0.000001cm}
\subfigure[]{\includegraphics[width=8cm,height=4.5cm]{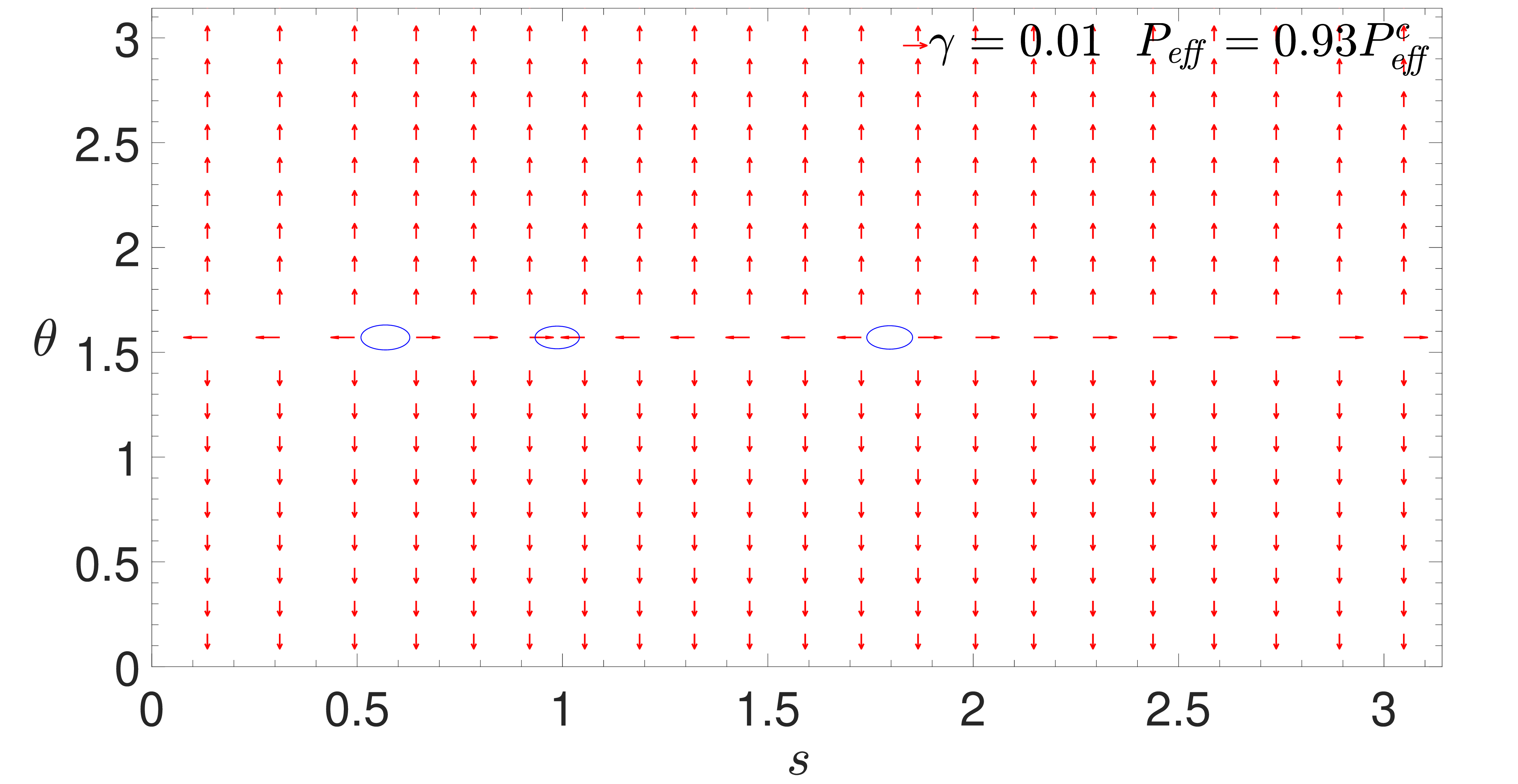}\label{Fig8b}}~
\\
\subfigure[]{\includegraphics[width=8cm,height=4.5cm]{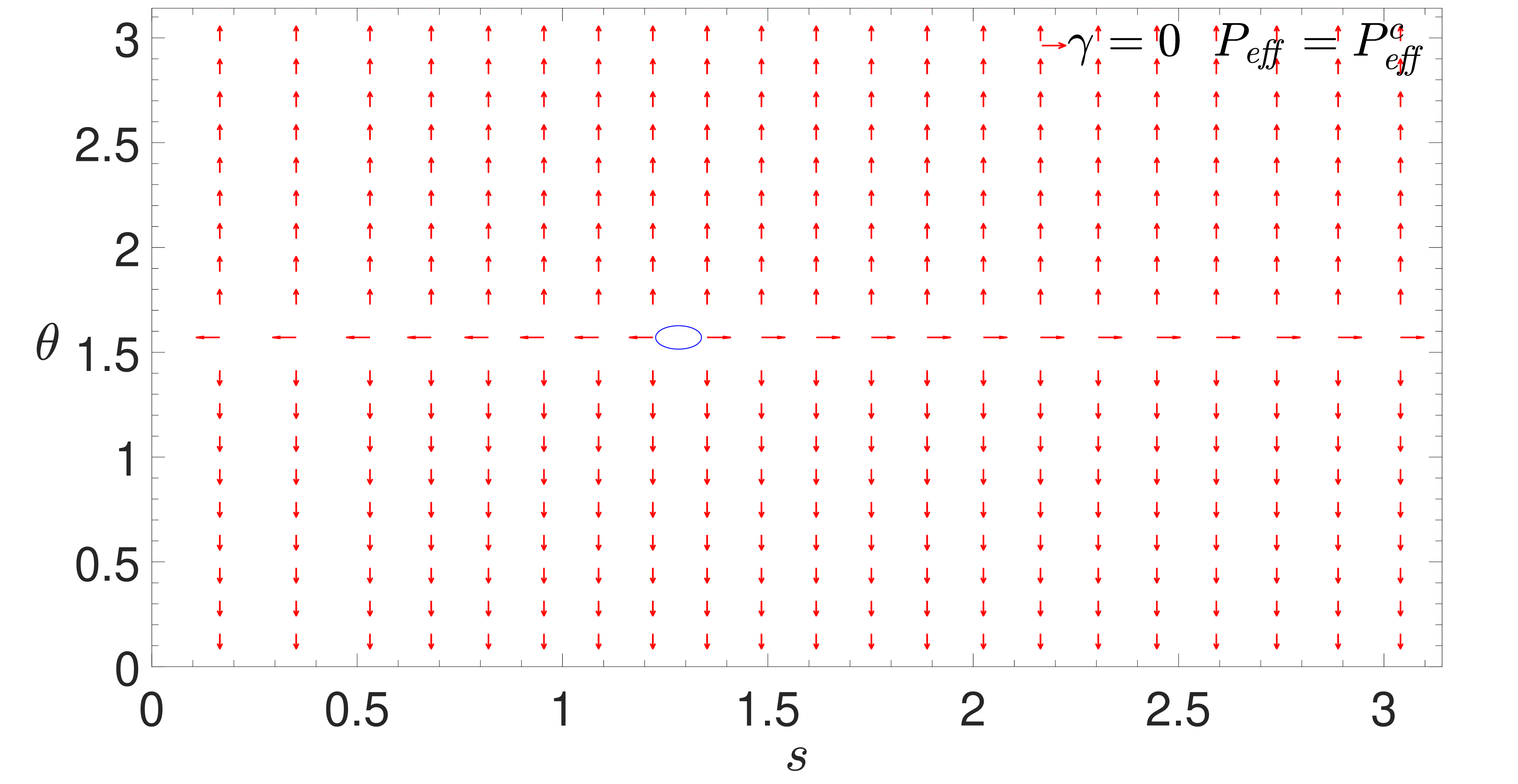}\label{Fig8c}}~
\hspace{0.000001cm}
\subfigure[]{\includegraphics[width=8cm,height=4.5cm]{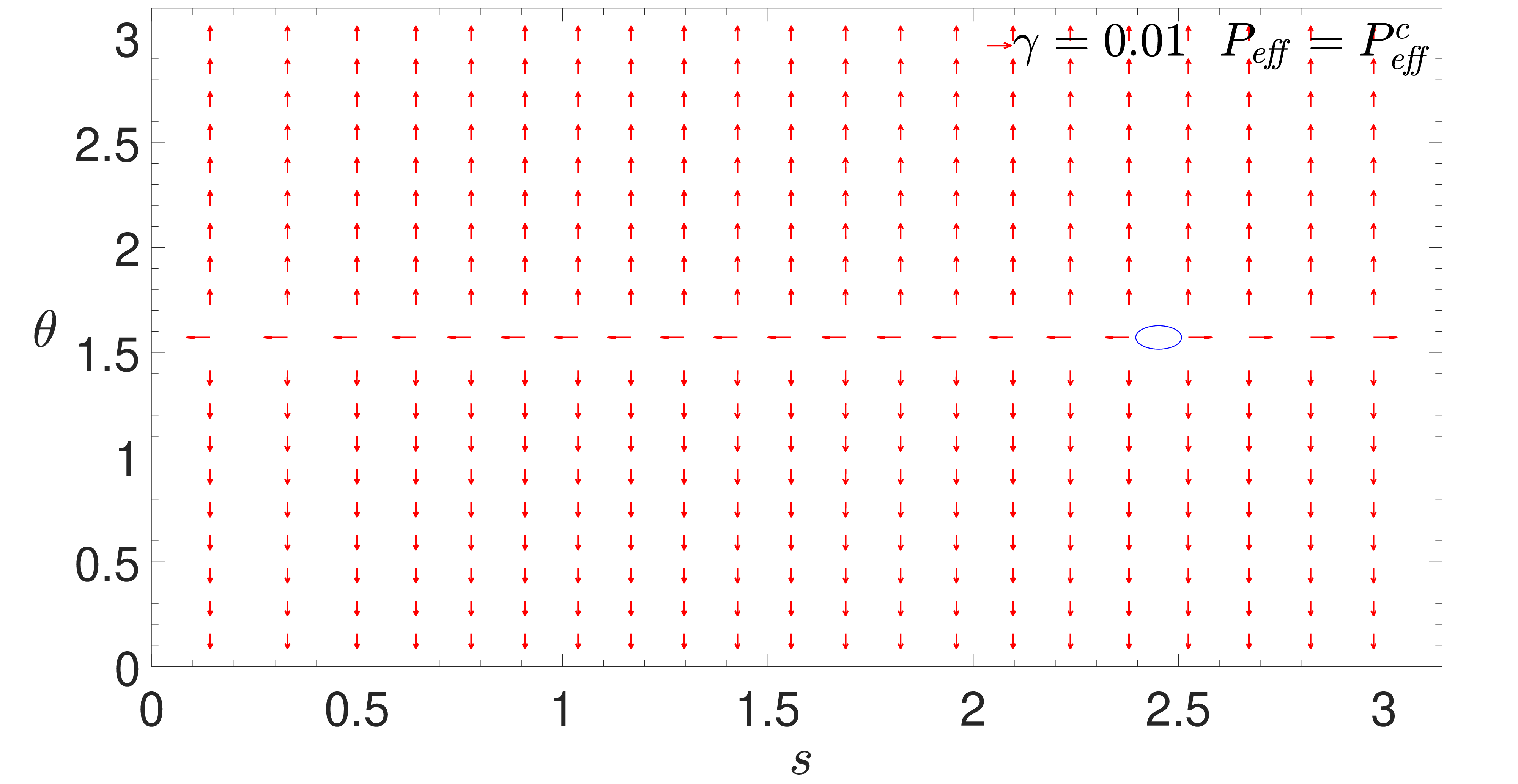}\label{Fig8d}}~
\\
\subfigure[]{\includegraphics[width=8cm,height=4.5cm]{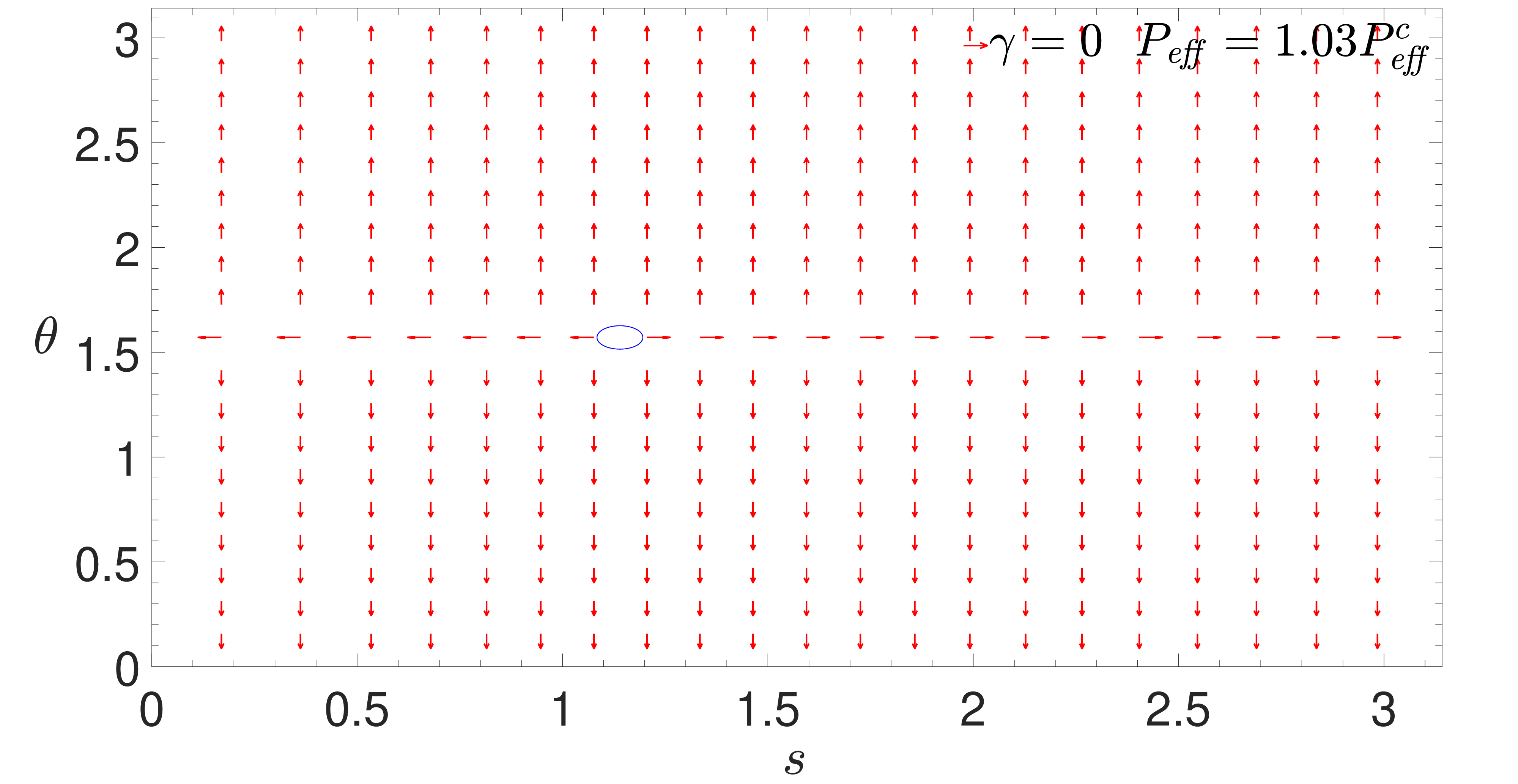}\label{Fig8c}}~
\hspace{0.000001cm}
\subfigure[]{\includegraphics[width=8cm,height=4.5cm]{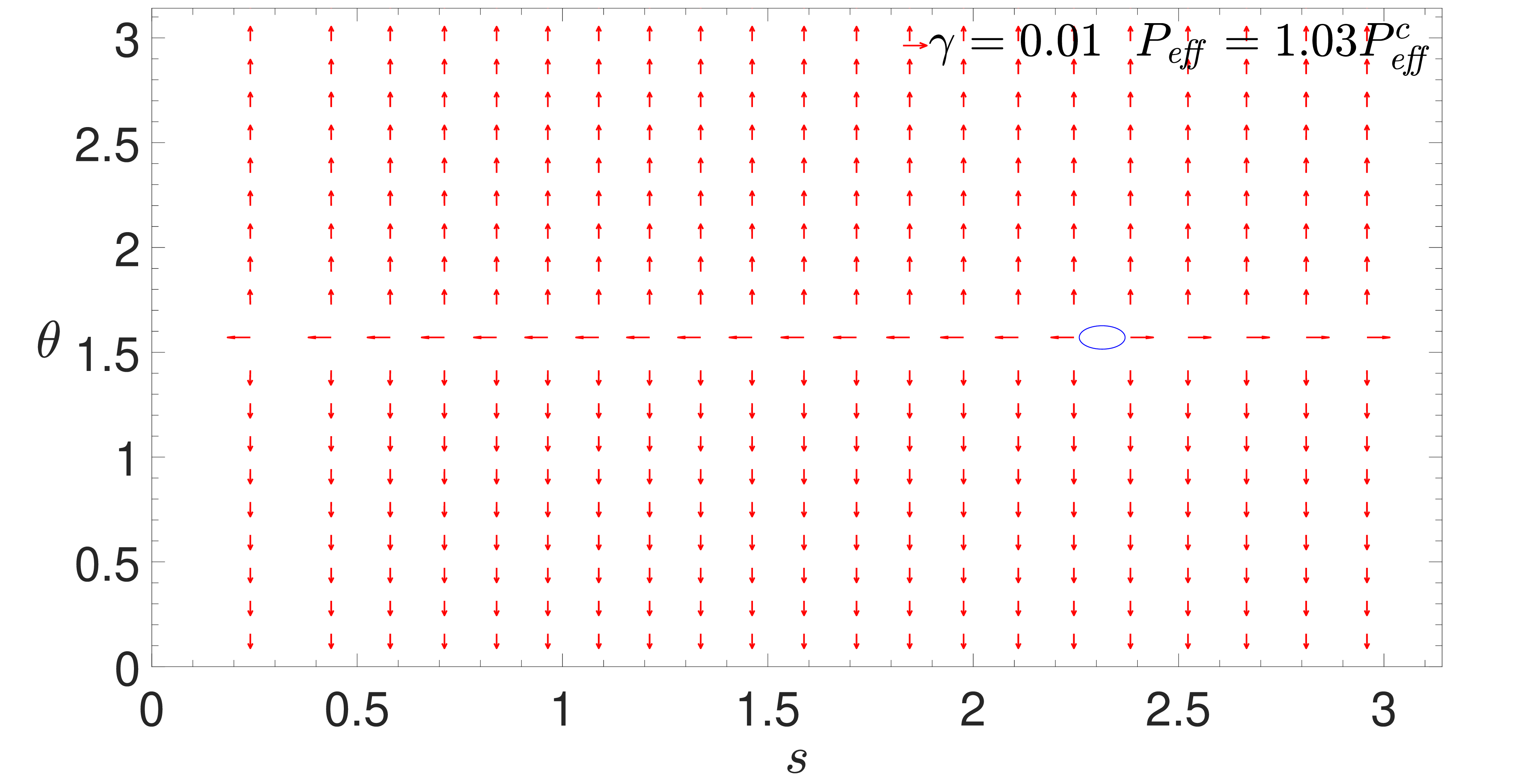}\label{Fig8d}}~
\vskip -1mm \caption{(Color online) The vector field diagrams on a segment of the ($S-\theta$) plane for different values of ${{P}_{eff}}$ with the charge $Q=1$ and fixed nonlinear parameter $\gamma=0,0.01$.}
\end{figure}

When ${{P}_{eff}}<P_{eff}^{c}$, we plotted the curve for $\gamma=0$ and $\gamma=0.01$, as shown in FIG 4.1 (a)and (b), the three blue loops represent the topological charge $\omega=(+1, -1, +1)$ and the total topological number $W =+1$. While for ${{P}_{eff}}\ge P_{eff}^{c}$, in FIG 4.1(c), (d) and(e), (f), we also plotted the curve for $\gamma=0$ and $\gamma=0.01$ respectively. The single blue loop represent the one topological charge $\omega=+1$, so that the total topological number is $W =+1$. Additionally, this indicates that the topological number remains invariant under variations of the nonlinear parameter $\gamma$, that is, its value does not affect the total topological number of the equivalent thermodynamic system.

\section{Discussions and Conclusions}\label{five}

We explore the thermodynamic characteristics of two-horizon coexistence region from different angles, demonstrating that the equivalent thermodynamic system exhibits phase transition features similar to those observed in AdS black holes. Furthermore, the influence of the nonlinear parameter $\gamma$ on these phase transitions is discussed. These findings provide a deeper understanding of dS spacetime, establish a novel pathway for investigating its thermodynamic properties, and offer new perspectives for revealing the nature of microscopic particles within black holes.

In Sec.4, we apply the topological approach, recently developed for studying black holes, to investigate the equivalent thermodynamic system within the two-horizon coexistence region of dS spacetime. We find that when treating the equivalent thermodynamic system as a canonical ensemble, it exhibits topological properties analogous to those of AdS black holes, characterized by a total topological number $W=+1$. Crucially, the nonlinear parameter $\gamma$ in the spacetime metric is found to have no effect on this topological number. Furthermore, the stability and instability within the coexistence region are determined by analyzing the system's positive and negative winding numbers.

Recent studies on the thermodynamics of AdS black holes have revealed that all black holes can be classified into four universal thermodynamic classes, denoted as ${{W}^{1-}}$, ${{W}^{0+}}$, ${{W}^{0-}}$ and ${{W}^{1+}}$ \cite{20,51,57,58,59,60}. These findings uncover a universal topological classification underlying black hole thermodynamics, providing fundamental insights into the basic principles of quantum gravity. Moreover, for dS spacetime featuring a coexistence region of two horizons, it remains to be rigorously established whether the thermodynamic system described by equivalent thermodynamic state parameters satisfies this universal classification.

\section*{Acknowledgments}
We would like to thank Prof. Ren Zhao for his indispensable discussions and comments. This work was supported by the Natural Science Foundation of Shanxi Province (Grant Nos. 202303021211180, 202203021221211) and the Program of State Key Laboratory of Quantum Optics and Quantum Optics Devices (KF202403).

\end{document}